\documentclass[usenatbib]{mn2e}

\usepackage{graphicx}

\usepackage{amssymb}
\usepackage{amsmath}
\usepackage{color}

\usepackage[colorlinks,urlcolor=blue,citecolor=blue,linkcolor=blue]{hyperref}

\usepackage{natbib}
\usepackage{wrapfig}

\newcommand{\be}{\begin{equation}}
\newcommand{\ee}{\end{equation}}

\newcommand{\ba}{\begin{eqnarray}}
\newcommand{\ea}{\end{eqnarray}}

\newcommand{\psrA}{{PSR~J1101--6101}}
\newcommand{\psrB}{{PSR~B2224+65}}
\newcommand{\psrC}{{PSR~J1509--5850}}
\newcommand{\psrD}{{PSR~B0355+54}}
\newcommand{\psrE}{{PSR~J2021+3651}}

\renewcommand{\v}{{\bf v}}

\newcommand\cf{it{cf.}}

\newcommand\eg{\textit{e.g.}}

\newcommand{\Bf}{{magnetic field}}
\newcommand{\Bfs}{{magnetic fields}}

\newcommand{\Lf}{Lorentz factor}

\newcommand{\apj}{ApJ}
\newcommand{\apjs}{ApJS}
\newcommand{\apjl}{ApJ}
\newcommand{\mnras}{MNRAS}

\newcommand{\aap}{A\&A}

\newcommand{\araa}{ARA\&A}

\newcommand{\ssr}{SSRv}
\newcommand{\planss}{P\&SS}

\title[Kinetic ``jets'' from rapid pulsars]{Kinetic ``jets'' from fast moving pulsars}
\author[Barkov et al.]{Maxim V. Barkov$^{1,2}$\thanks{Correspondence author: mbarkov@purdue.edu (MVB)},
Maxim Lyutikov$^{1}$, Noel Klingler$^{3}$ and Pol Bordas$^{4}$.
 \\
$^{1}$ Department of Physics and Astronomy, Purdue University, West Lafayette, IN 47907-2036, USA\\
$^{2}$ Astrophysical Big Bang Laboratory, RIKEN, 351-0198 Saitama, Japan\\ 
$^{3}$ Department of Physics, The George Washington University, Washington, DC 20052, USA \\
$^{4}$ Departament de F\'{i}sica Qu\`{a}ntica i Astrof\'{i}sica, Institut de Ci\`{e}ncies del Cosmos (ICCUB), 
\\
Universitat de Barcelona, IEEC-UB, Mart\'{i} i Franqu\`{e}s 1, E08028 Barcelona, Spain}

\begin{document}
\date{Received/Accepted}
\maketitle

\begin{abstract}
Some fast-moving pulsars, like the Guitar and the Lighthouse, exhibit asymmetric non-thermal emission features that extend well beyond their ram pressure confined pulsar wind nebulae (PWNe). Using 3D relativistic simulations we explain these features as kinetically streaming pulsar wind particles that escaped into the interstellar medium (ISM) due to reconnection between the PWN and ISM magnetic fields. The structure of the reconnecting magnetic fields at the incoming and outgoing regions produce highly asymmetric magnetic bottles, and result in  asymmetric extended features. For the features to become visible, the ISM magnetic field should be sufficiently high,  $B_{\rm ISM}>10$~$\mu$G.
We also discuss archival observations of PWNe displaying evidence of kinetic jets: the Dragonfly PWN (PSR J2021+3651), G327.1--1.1, and MSH 11--62, the latter two of which exhibit ``snail eyes'' morphologies. We suggest that in those cases the pulsar is moving along the ambient magnetic field in a frisbee-type configuration.


\end{abstract}


\section{Introduction}
\label{intro}

Rotation powered pulsars produce relativistic winds that interact with the surrounding medium  to form  pulsar wind nebulae (PWNe). Strong shocks at the wind/medium interface lead to efficient particle acceleration and radiation of non-thermal emission (see, e.g., \citealt{1974MNRAS.167....1R,1984ApJ...283..694K,2006ARA&A..44...17G} for reviews). 
When a pulsar travels through the interstellar medium (ISM) at supersonic velocities, the structure of the system is strongly affected, and the morphological and spectral properties of these high-velocity PWNe depart significantly from their more common low-velocity analogues. The ram pressure exerted by the oncoming ISM confines the pulsar wind to the direction opposite that of the pulsar's motion, forming a bow shock compact nebula (CN) and an extended pulsar tail.  Best observed in X-rays, where high-resolution morphological studies are technologically feasible with the current space-based facilities, about 30 of such fast-moving PWNe have been imaged to date \citep[see e.g.][]{kp08,kpkr17}. 

The typical size of a ram pressure-confined PWN can be taken as the stand-off distance $r_s$ of the bow shock apex, where the pressure of the pulsar wind, $\dot{E}/( 4 \pi r_s^2 c)$, is balanced by the ram pressure of the ISM, $\rho_{\rm ISM} v_p^2$.
\be
r_s = \sqrt{ \frac{ \dot{E}}{ 4 \pi c \mu m_p n_{\rm ISM} v_p^2}} \approx 4\times 10^{-3} \dot{E}^{1/2}_{36} v^{-1}_{p,8} n_0^{1/2}{\rm pc}
\label{eq:rs}
\ee
where $\dot{E}$ is the pulsar's spin-down luminosity, $n_{\rm ISM}$ is the ISM number density, $v_{p,8} = v_p/10^8$ cm~s$^{-1}$ is the pulsar velocity, $m_p$ is proton mass and $\mu$ is ISM specific weight.  The numerical value for {the} luminosity can be scaled to $\dot{E}_{36}= \dot{E}/10^{36}$ erg s$^{-1}$ and the ISM number density to $n_0 =1$ particle cm$^{-3}$. In the tail, the PWN can extend well-beyond the length scale of the stand-off distance (e.g., up to parsec-scale distances), yet in the head of the CN, [\ref{eq:rs}] provides an estimate of the expected forward extent of the nebula.

\begin{figure}
\includegraphics[width=0.97\linewidth]{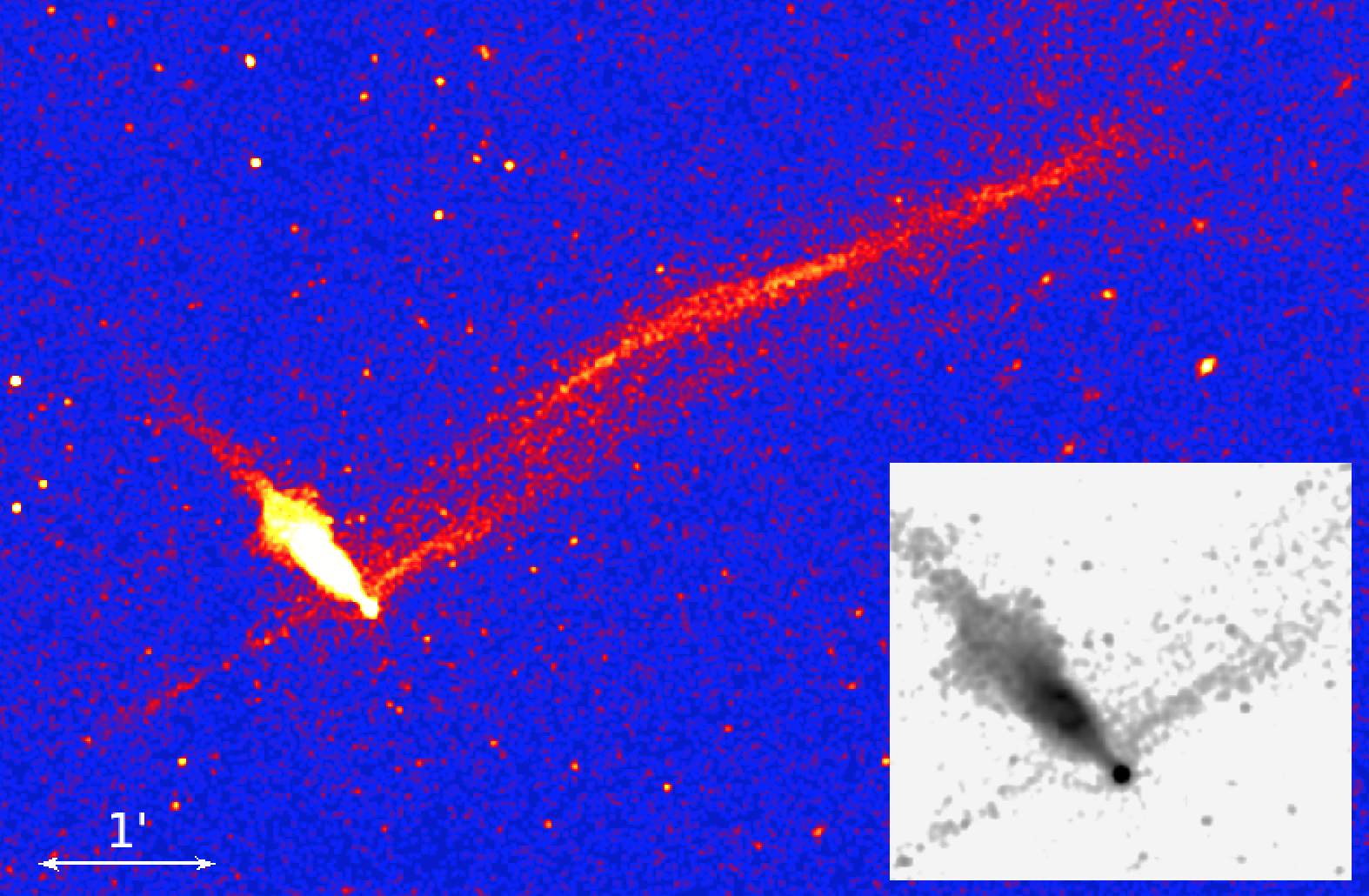}
\caption{{\sl Chandra} images showing the extended emission features in the Lighthouse PWN powered by \psrA{} \citep{ppba16}, on large and small (inset) scales.}
\label{fig:lighthouse}
\end{figure}

Contrary to the hydrodynamical expectation that pulsar wind should be contained within the boundaries of the contact discontinuity that separates the wind from the ISM,
 observations of some high-velocity PWNe have revealed the presence of elongated  features extending well beyond the stand-off distance $r_s$ in the forward or sideways directions; see, e.g., the cases of the Lighthouse PWN of \psrA\ \citep[][ Fig. \ref{fig:lighthouse}]{ppba16}, the Guitar PWN of \psrB\ \citep{2003IAUS..214..135W,2007A&A...467.1209H, 2010MNRAS.408.1216J}, 
\psrC{} \citep{2016ApJ...828...70K} and the Mushroom PWN of \psrD{} \citep{2016ApJ...833..253K}. These features are misaligned with respect to the pulsars' proper motion directions and extend rather deep into the medium, far outside the bow shock boundary, reaching parsec-scale lengths in some instances. 

These so-called ``misaligned outflows'' are sometimes interpreted as jets. However, they {cannot} be jets in the sense of a confined hydrodynamic flow, as any such flow {cannot} propagate much beyond the stand-off distance (\ref{eq:rs}).  
Also, in two of the known cases, J1509--5850 and B0355+54, both pulsar jets and counter-jets have been resolved and are confined within the bow shock, thus ruling out this explanation. 
To stress this difference, in this paper we will refer to these misaligned outflows as ``kinetic jets''. As we demonstrate in this study, these features can be produced by the kinetic streaming of high-energy non-thermal (NT) particles that escape the shocked pulsar wind near the head of the bow shock nebula, a scenario proposed by \cite{ban08} to explain the linear feature observed in the Guitar Nebula. 

Importantly, in nearly all cases, these extended features  show a remarkable asymmetry with respect to the PWN axis, displaying emission much more extended in one direction than in the other -- sometimes interpreted as a ``counter-jet'' (see, e.g., \citealp{2014A&A...562A.122P, ppba16}). Furthermore, these ``double-jet'' features can display significantly differing surface brightnesses, often attributed to the Doppler boosting of synchrotron emission from these collimated outflows in opposite directions.  
{However, the spectral properties of these jet-like features are difficult to explain in a ballistic jet scenario for some systems 
\citep{ppba16,  2010MNRAS.408.1216J}, for which alternative mechanisms (e.g. particle re-acceleration along the jets) may be invoked \citep[see ][]{2017SSRv..207..235B}. 
}

{\cite{ban08} proposed a scenario in which} the long jet-like X-ray feature observed in Guitar Nebula 
arises due to the leakage of high energy electrons accelerated at the bow shock region. When the gyroradii of high energy particles exceeds the bow shock stand-off distance, they can no longer be contained within the bow shock and escape,
crossing the contact discontinuity and 
gyrating along the ISM magnetic field.  This scenario explains the apparent linear morphology and seemingly-random offset from the pulsar velocity of these features (which trace the ISM field lines) as well as their relatively large sizes. 

In {this} paper we further develop this model and confirm the properties of extended jet-like features expected in that scenario through dedicated 3D numerical simulations of  ``kinetic jets'' emanating from fast-moving pulsars, assuming a range of conditions for the system and surrounding ISM properties. The setup of these numerical simulations is outlined in Sect.~\ref{s:ss}, and the obtained results are reported in Sect.~\ref{s:res}, where it is shown that ISM magnetic field lines reconnect with pulsar wind magnetic field lines, leading to an asymmetrical configuration for NT particle escape. The applications of our simulations to some fast-moving PWNe displaying extended X-ray jet-like features are discussed in {Sect.~\ref{s:applications}}. The final conclusions of this study are summarized in Sect.~\ref{s:conc}.



\section{Kinetic jets in bow shock PWN}
\label{s:disc}

\subsection{Overall properties}

About 30 runaway pulsars moving through the interstellar medium at supersonic velocities have been identified so far, based on observational properties obtained either in optical (e.g., through the detection of H$\alpha$ emission lines produced around their bow-shocks) or via the synchrotron emission seen in radio and/or X-rays (see, e.g., \citealt{kpkr17} for a review). 
%
%

\begin{figure*}
\includegraphics[width=1\linewidth]{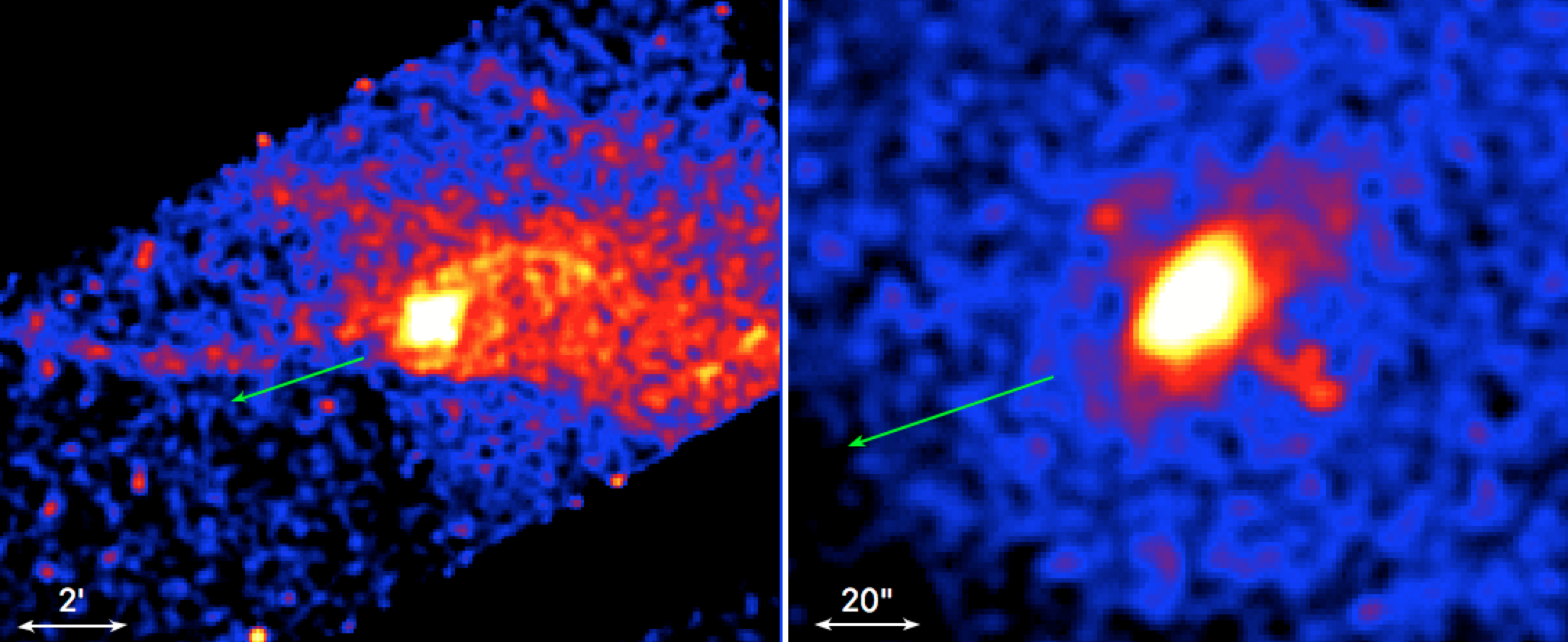}
\caption{The {\sl Chandra} image 
of \psrE{} shows a bow shock head (left panel) and clearly resolved torus+jets within the bow shock (with the equatorial outflow dominating; right panel).  The torus is viewed at angle $\sim 80^\circ$ \citep{2008ApJ...673..411N}.  J2021 has a velocity/spin offset of $\sim 45^\circ$ (as we have in our models, see the Y projection in Fig.~\ref{fig:Ebd}).  
Faint outflow-like emission is seen ahead of the bow shock, which is likely a kinetic jet.  The green arrows show the inferred direction of pulsar motion.}
\label{fig:j2021}
\end{figure*}

In the four previously-reported instances of kinetic jets mentioned above,
the pulsars move at velocities ranging from about $v=61$ km s$^{-1}$ ($\mu=12.3\pm0.4$ mas yr$^{-1}$ for B0355+54, \citealt{chatterjee2004}) up to $v_{\rm PSR} \gtrsim 1000$~km s$^{-1}$ (the highest known pulsar velocities), inferred from the measurements of (or an upper limits on) $r_s$ for PSR J1101--6101 (see Eq.\ \ref{eq:rs}), and inferred from proper motion measurements for B2224+65 ($\mu=182\pm3$ mas yr$^{-1}$, though the distance is not reliably known; \citealt{cc2004}). As typical ISM sound speeds range from $c_{\rm ISM}\sim1$ km s$^{-1}$ (the ``cold'' phase) to a few times 10 km s$^{-1}$ (the ``warm'' phase; see \citealt{cox2005}), these SPWNe are highly supersonic, as is also shown by their morphologies. 
The pulsar characteristic ages span from about $10^5$ yr, considered ``middle-aged'' for PWNe, to $10^6$ yr for B2224+65, which is the oldest pulsar known with an X-ray PWN\footnote{Note that in the Guitar Nebula, only an H$\alpha$ bow shock and the kinetic jet are present; no synchrotron tail is seen, unlike the other instances.}. Their spin-down powers range from 
$\dot{E} =1.3\times 10^{36}$~erg s$^{-1}$ (PSR J1101--6101), typical of young to middle-aged energetic pulsars, to 
$\dot{E} = 1.2\times10^{33}$~erg s$^{-1}$ (PSR B2224+65), the least-energetic pulsar with an X-ray nebula.  Total X-ray luminosities for these sources lie between $L_{\rm X} \sim 10^{30}$~erg s$^{-1}$ to $10^{33}$~erg s$^{-1}$ (see \citealt{kp08} and references therein).  Compared to the larger population of X-ray PWNe, the Guitar is an outlier both in its low spin-down power, old age, and extremely high velocity.  

Kinetic jets from SPWN can show lengths from a few to more than 10~pc (e.g., in the Lighthouse Nebula), and display an orientation with respect to the pulsar proper motion (misalignement) in the range of $\sim 33^{\circ}$ (e.g., in J1509-5850; \citealt{2016ApJ...828...70K}), to $\sim 118^{\circ}$ (in the Guitar Nebula; \citealt{2012ApJ...747...74H}). The outflows are remarkably narrow, with width-to-length ratios of $\sim 0.15$ to $0.20$, and can have counter-jets extending along the same direction symmetrically with respect to pulsar position, but displaying a much lower luminosity. The width, on the other hand, can be attributed {in the kinetic jet scenario studied here} to the diffusion of high-energy particles into several quasi-parallel magnetic field lines of the surrounding medium.  

The X-ray luminosity of kinetic jets can range from $\sim 1/3$ of the total X-ray luminosity of the PWN (e.g., in the Lighthouse Nebula; \citealt{2014A&A...562A.122P}; and J1509) to 0.07 (in B0355). Their spectra do not show significant changes either along the main axis nor in the transverse direction\footnote{The J1509 outflow shows possible evidence of spectral cooling, as $\Gamma$ changes from $1.50\pm0.20$ to $1.98\pm0.11$ between the near- and far-halves of the outflow, however, the outflow's combined spectrum is identical to that of the compact nebula within the bow shock, $\Gamma\approx1.80\pm0.10$.}, implying that cooling timescales (e.g., via synchrotron emission) are longer than the particles' propagation timescale along across the jet. Kinetic jets display relatively hard spectra, $\Gamma_{\rm Guitar}\approx 1.3\pm0.1$, $\Gamma_{\rm Lighthouse}=1.7\pm0.1$, $\Gamma_{\rm J1509}=1.81\pm0.10$, and $\Gamma_{\rm B0355}=1.6\pm0.3$.  In all cases (except the Guitar, which does not have a detectable synchrotron bow shock compact nebula; CN), the spectra of the kinetic jets are approximately the same as the spectra of the CNe from which they originate.  It is also worth noting that none of the kinetic jets are seen in radio, even when their PWNe are (e.g., J1509 and the Lighthouse).  This is likely because the magnetic bottling effects that occur near the reconnection zones suppress the escaping of low energy particles that produce the radio emission.

Additionally, evidence of another kinetic jet has been reported in the PWN of the old ($\tau=1.2$ Myr) radio-quiet, $\gamma$-ray pulsar J2055+2539 ($\dot{E}=5\times10^{33}$ erg s$^{-1}$) by \citet{marelli2016}. This PWN features highly collimated $12'$ and $4'$ outflows, interpreted as a kinetic jet and synchrotron tail (though it is not clear which is which, as the proper motion of the pulsar is not yet known).  Although both extensions feature spectra typical of kinetic jets and tails, $\Gamma=1.82\pm0.08$ and $1.62\pm0.2$, their spectra can also be described by thermal Bremsstrahlung models.  This is resemblant of PSR J0357+3205 (a.k.a.\ Morla; \citealt{marelli2013}), which also features a collimated outflow whose spectrum can be adequately described by either a PL or Bremsstrahlung model, but which is aligned with the direction of pulsar motion yet puzzlingly ``disconnected'' from the pulsar by $\sim$30$''$.


\subsection{Kinetic Jets of the Dragonfly PWN (PSR J2021+3651)}

\begin{figure*}
\includegraphics[width=1\linewidth]{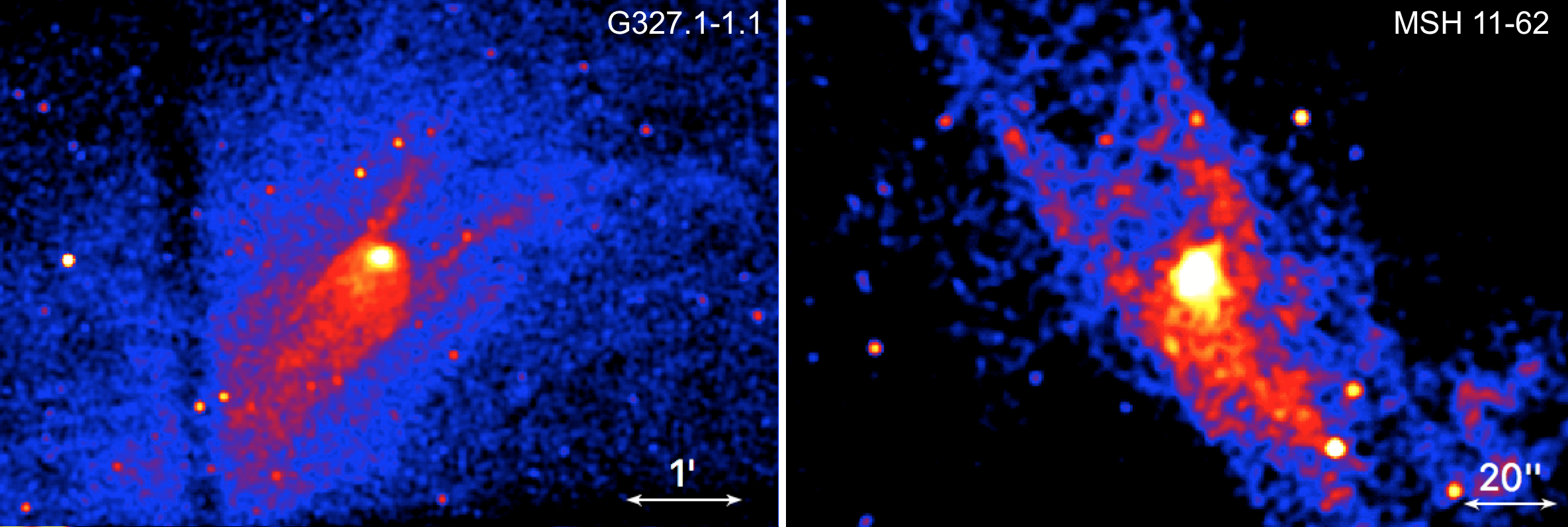}
\caption{{\sl Chandra} images of PWNe inside SNRs G327.1--1.1 and MSH 11--62 showing  kinetic jets with what we call ``snail eyes'' morphology. See Fig. \protect\ref{fig:snail_sk} for qualitative explanation.}
\label{fig:SNR-PWNe}
\end{figure*}

PSR J2021+3651, producing the Dragonfly PWN, is a young ($\tau_c=P/(2\dot{P})=17$ kyr), energetic ($\dot{E}=3.4\times10^{36}$ erg s$^{-1}$), Vela-like pulsar with period $P=103.7$ ms.  Its  dispersion measure (DM) of 369 pc cm$^{-3}$ suggests a distance $d\sim10.5$ kpc (using the \citealt{ymn17} electron density model), however, after {accounting for} deep {\sl Chandra} observations, \citet{vrn08} argue its estimated distance is more likely in the 3-4 kpc range.  Although its velocity is not known, the PWN clearly displays a bow shock head and a small pulsar tail (see Fig.~\ref{fig:j2021}).  The large projected standoff distance of the bow shock apex (43$''$; 0.8 pc at 4 kpc) and the fact that the torus and jets are not  deformed by the ram pressure suggest that the pulsar is only mildly supersonic.  Indeed, estimating the pulsar velocity from balancing the pulsar wind and ram pressure gives $v=(\dot{E}/4\pi c m n r^2)^{1/2}\sim10$ km s$^{-1}$, suggesting the pulsar is mildly supersonic with the ISM in the cold phase (in which the sound speed $c_s\sim$ a few km s$^{-1}$) for an assumed typical hydrogen number density $n_H=1$ cm$^{-3}$ (where $m$ is the hydrogen mass).  Note that the above estimate assumes {an isotropic} wind, which is not the case here; thus the actual velocity is likely higher.  The orientation of the torus with respect to the pulsar's direction of motion suggests it is moving 
with a velocity vector offset from its spin axis by $\sim45^\circ$,
making it a useful reference with which to compare our simulations of the slow case ({see below}).  

An extended narrow structure is seen protruding from the bow shock up to 7$'$ ahead of the pulsar (see Fig.~\ref{fig:j2021}).  \citet{vrn08} report that the spectrum of the so-called ``arc'' structure {is best fit with} an absorbed power-law (PL) model with slope $\Gamma=1.66\pm0.25$ -- a hard spectrum which is typical of kinetic jets. 

\subsection{G327.1--1.1 and MSH 11--62: ``Snail eyes'' Morphology}


Two young PWNe residing in their host SNRs have been observed to feature prong-like outflows (morphologically resembling ``snail eyes'') ahead of their bow shocks, oriented parallel to the direction of pulsar motion: the ``Snail PWN'' in G327.1--1.1 \citep{temim09} and MSH 11--62 (G291.0--0.1; \citealt{slane12}), see Fig.~\ref{fig:SNR-PWNe}.  These structures exhibit PL spectra typical of kinetic jets, with $\Gamma\approx1.8$ and $\approx1.4$, respectively (see Figures 1 and 3 of \citealt{kkcp17}, and \citealt{temim2015}).  
The proper motion of these pulsars are not known, {although} simulations by \citet{temim2015} suggest a typical velocity $v\sim400$ km s$^{-1}$ for the pulsar producing the snail PWN\footnote{No pulsations have been detected from {either of these two pulsars}.}. 
Compared to the other instances of kinetic jets, the pairs of jets in these examples appear approximately symmetric both in structure and surface brightness.   

\section{Formation of kinetic jets}

\subsection{Reconnection between ISM and PWN magnetic fields}

NT particles are confined by magnetic fields in the direction perpendicular to the field, but  can travel large distances along these lines (parallel to the field). As a pulsar moves through the ISM, the interstellar \Bf\ lines become draped around the PWN. Even if the \Bf\ is at  sub-equipartition  in the bulk flow, a narrow draping layer with near-equipartition \Bf\ is created at the contact discontinuity \citep{1966PSS...14..223S,2006MNRAS.373...73L,2008ApJ...677..993D}. As a result, the contact discontinuity becomes a rotational discontinuity with  \Bfs\ of similar strength on both sides. Rotational discontinuities are prone to reconnection \citep[see, e.g.,][ and references therein]{2007MNRAS.374..415K,2016MNRAS.458.1939B}. The efficiency of reconnection at a given point on the contact/rotational discontinuity will depend on the relative orientation of the PWN and ISM \Bfs.

\begin{figure*}
\includegraphics[width=0.48\linewidth]{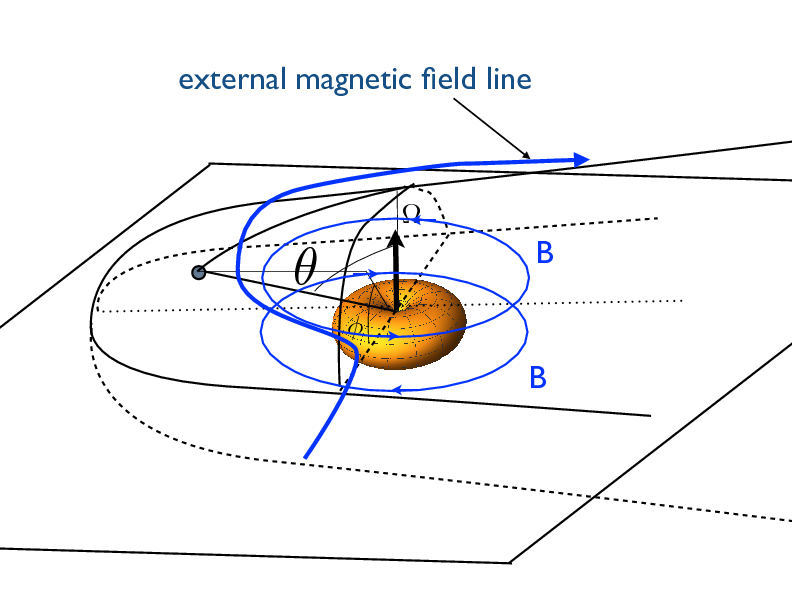}
\includegraphics[width=0.48\linewidth]{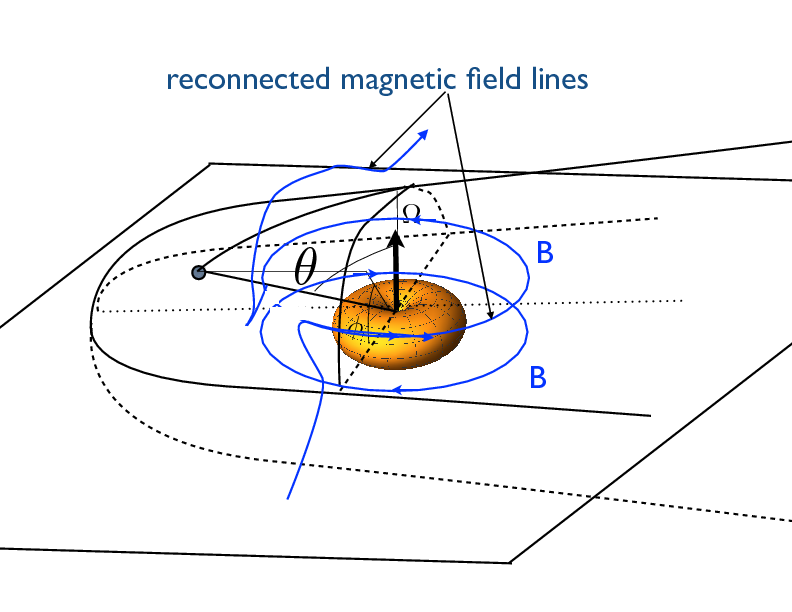}
\caption{Qualitative illustration of asymmetric magnetic bottling. A pulsar in the ``frisbee'' configuration (i.e., spin axis is oriented parallel to velocity direction) propagates 
through the ISM \Bf\ which is perpendicular to its direction of motion and spin axis. The internal \Bfs\ within the PWN are oriented in opposite directions in the northern and southern hemispheres. 
Reconnection between the draped external field and intrinsic fields creates back-and-forth asymmetric magnetic configurations. Full numerical simulations confirm this simple qualitative explanation, 
see Fig.~\ref{fig:sl}.}
\label{fig:sl0}
\end{figure*}

Reconnection between the pulsar wind and interstellar \Bf\ lines  will allow relativistic particles from the wind to escape into the ISM.
The number of NT particles that would escape in a given direction will depend on the fluctuations of the strength of the magnetic field along a field line -- regions of stronger magnetic fields will reflect some particles, forming a magnetic ``bottle''. 
As we demonstrate in this paper, particle escape can be highly asymmetric with respect to the incoming and outgoing parts of the \Bf\ lines due to the formation of these magnetic bottles.

\begin{figure}
\includegraphics[width=0.95\linewidth]{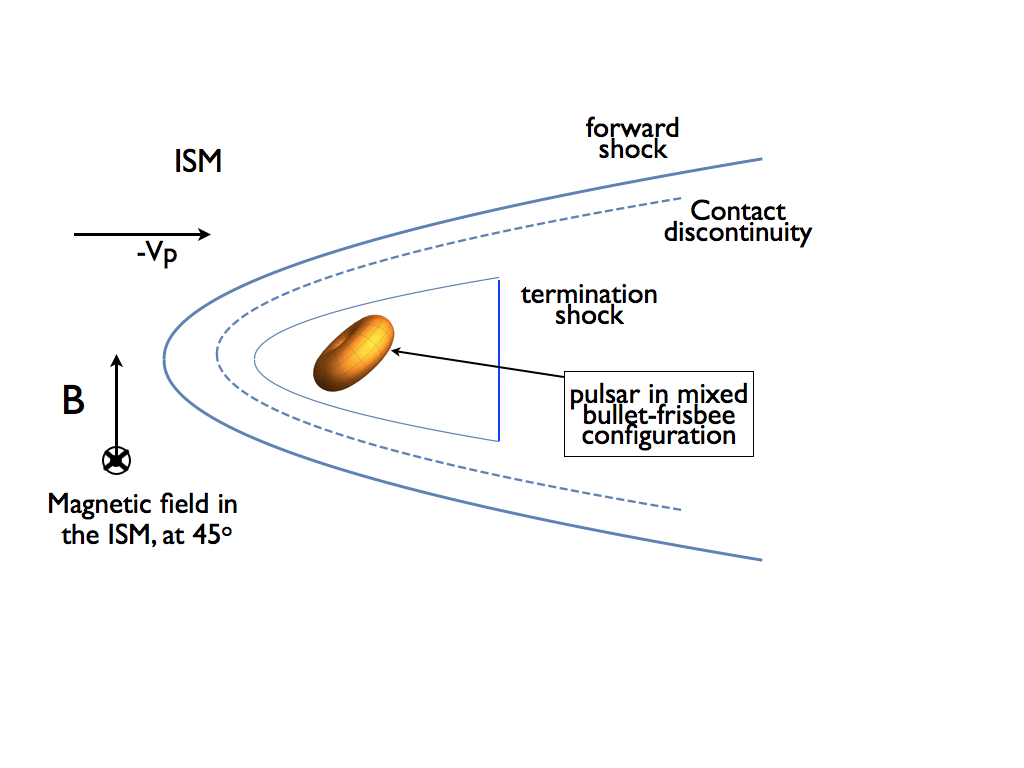}
\caption{Qualitative illustration of initial orientation of the pulsar relative to its motion in the ISM.}
\label{fig:op}
\end{figure}

\begin{table*}
\begin{tabular}{lccccccc}
\hline
\hline
  Coordinates         &  Left    &  $N_{\rm l}$   &   Left-center  &   $N_{\rm c}$ &   Right-center & $N_{\rm r}$ & Right  \\
\hline
&&&&&\\[-5pt]
 $ X  $     & \quad $-4$ & \quad $72$ & \quad $-1$ &\quad 144 & \quad $1$ &\quad 252 & \qquad $10$ \\
 $ Y  $     & \quad $-5$ & \quad $96$ & \quad $-1$ &\quad 144 & \quad $1$ &\quad 96  & \qquad $5$ \\
 $ Z  $     & \quad $-5$ & \quad $96$ & \quad $-1$ &\quad 144 & \quad $1$ &\quad 96  & \qquad $5$ \\
\hline
&&&&&\\[-5pt]
\end{tabular}
\caption{Parameters of the Grid}
\label{tab:grid}
\end{table*}

\begin{figure*}
\includegraphics[width=0.47\linewidth]{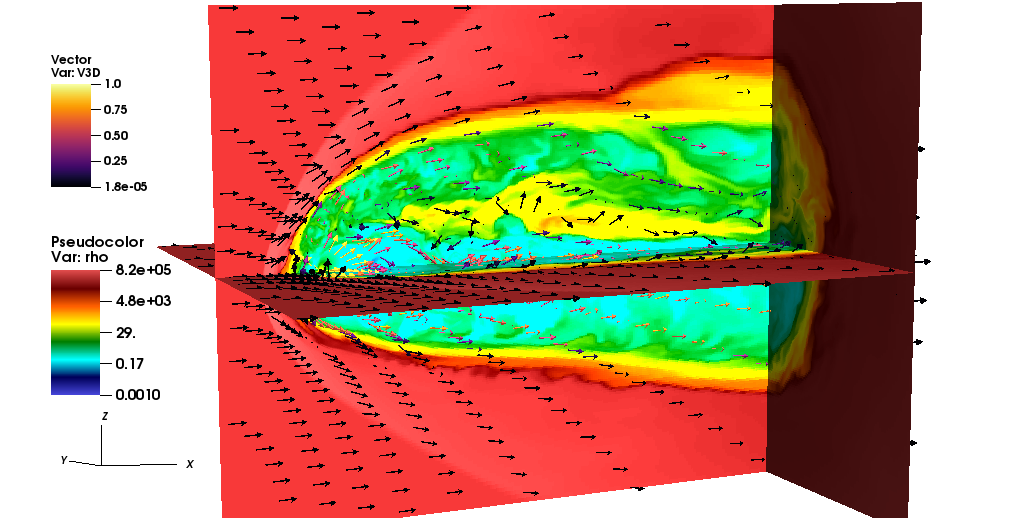}
\includegraphics[width=0.47\linewidth]{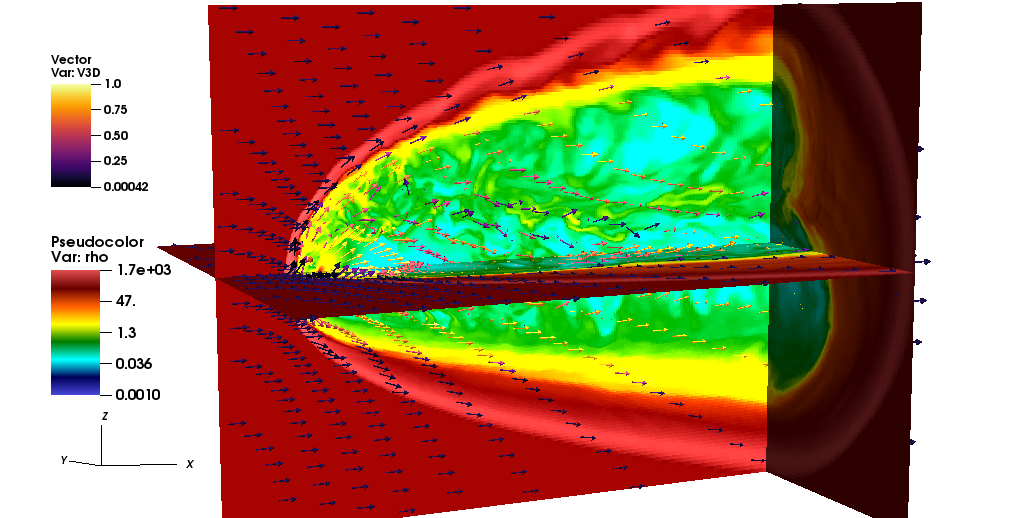}
\caption{Density cuts and velocity vectors for slow model (left) and fast model (right).}
\label{fig:rho}
\end{figure*}

\begin{figure*}
\includegraphics[width=0.47\linewidth]{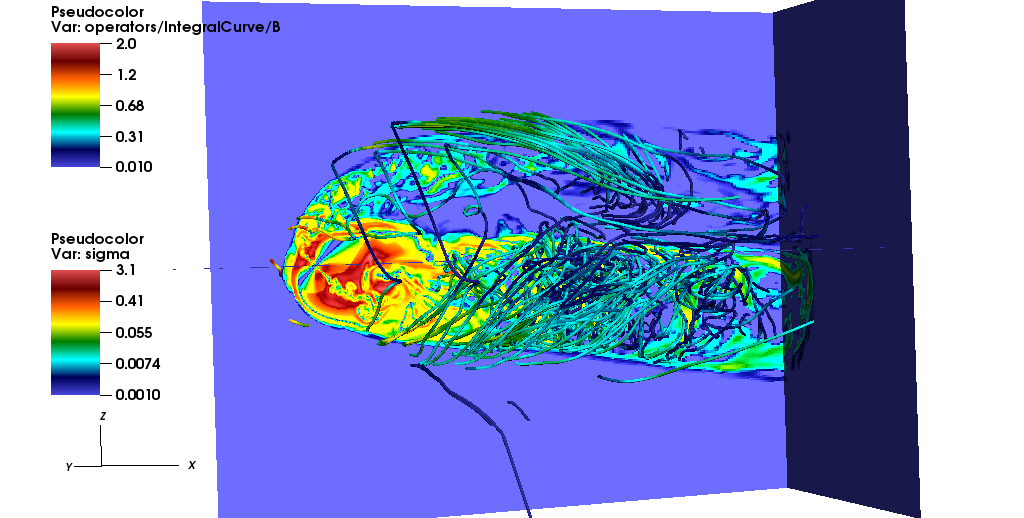}
\includegraphics[width=0.47\linewidth]{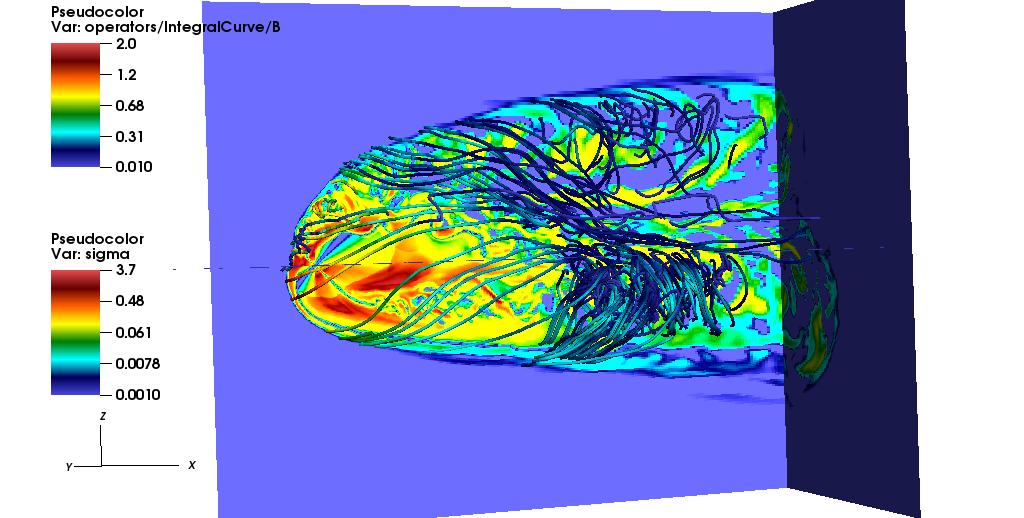}
\caption{Magnetization cuts and magnetic field lines for the slow model (left) and fast model (right).}
\label{fig:sigma}
\end{figure*}

\begin{figure*}
\includegraphics[width=0.47\linewidth]{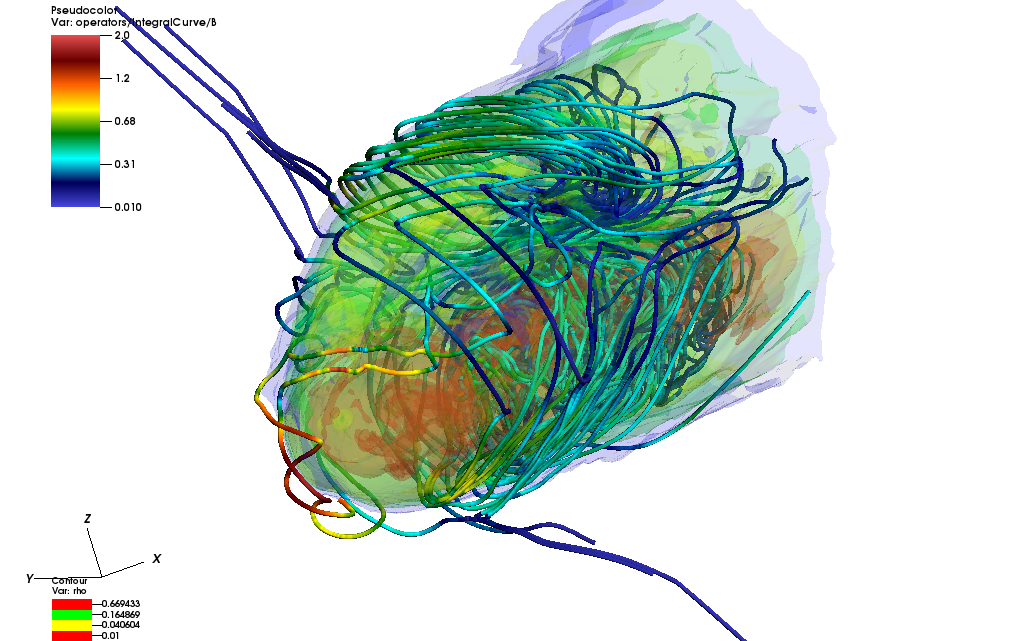}
\includegraphics[width=0.47\linewidth]{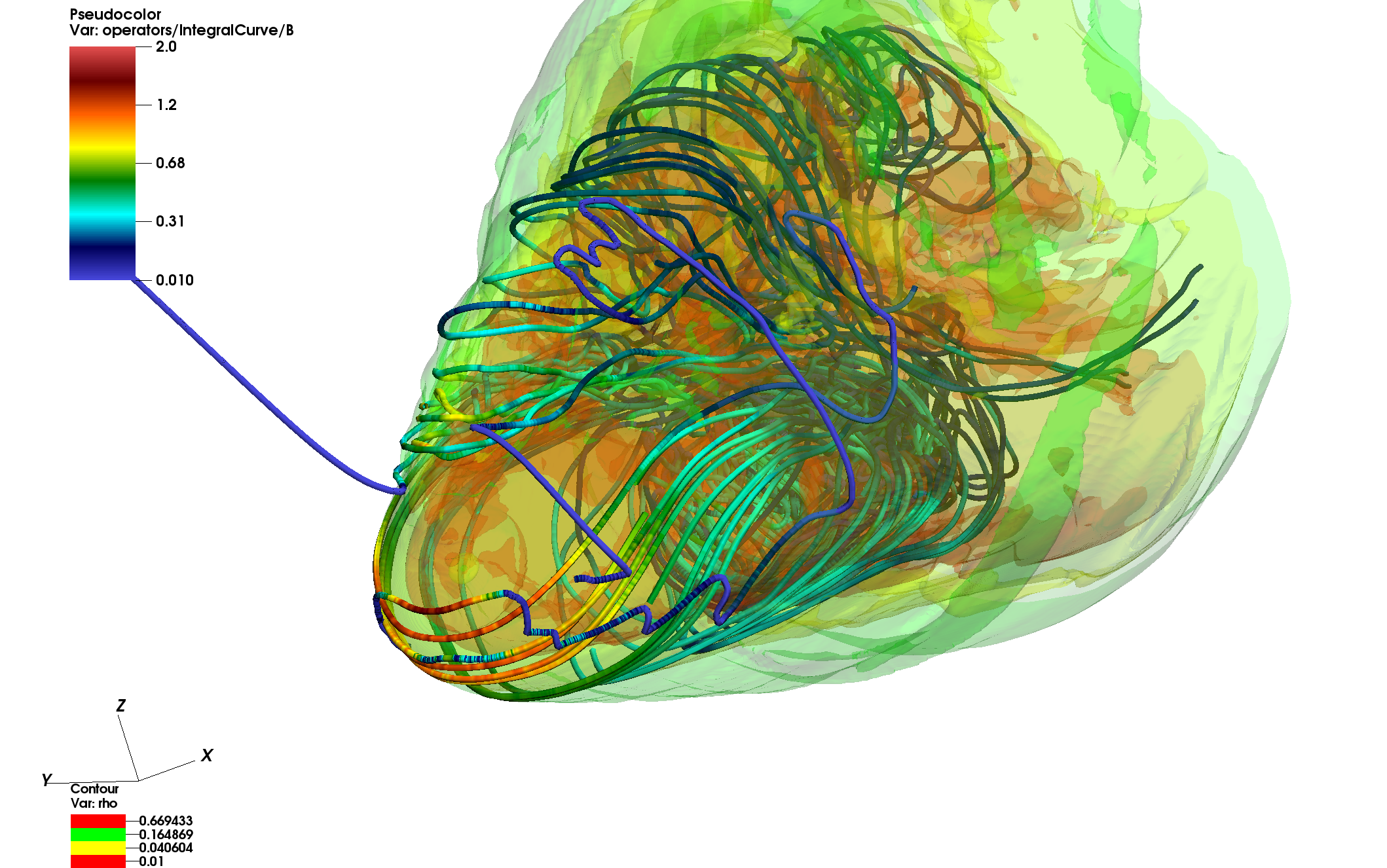}
\caption{Iso-density surfaces and velocity vectors for the slow model (left) and fast one (right).}
\label{fig:rhoB}
\end{figure*}

\subsection{Numerical Simulation Setup}
\label{s:ss}

The simulations were performed using a three dimensional (3D) geometry in Cartesian coordinates using the {\it PLUTO} 
code\footnote{Link http://plutocode.ph.unito.it/index.html} \citep{mbm07}. 
Spatial parabolic interpolation, a 3rd order Runge-Kutta approximation in time, and an HLL Riemann solver were used \citep{HLL83}. 
{\it PLUTO} is a modular Godunov-type code entirely written in C and intended mainly for astrophysical applications and high Mach number flows in multiple spatial dimensions. 
The simulations were performed on CFCA XC30 cluster of the National Astronomical Observatory of Japan (NAOJ).
The flow has been approximated as a relativistic gas of one particle species, and with the Taub equation of state. The size of the domain is $x \in [-4, 10]$, $y \mbox{ and } z \in [-5, 5]$. 
To maintain high resolution in the central region and along the tail zone we use a non-uniform resolution in the computational domain 
with the total number of cells $N_{\rm X} = 468$, 
and $N_{\rm Y} = N_{\rm Z}  = 336$.  More details are given in Table~\ref{tab:grid}.


\begin{figure*}
\includegraphics[width=85mm,angle=-0]{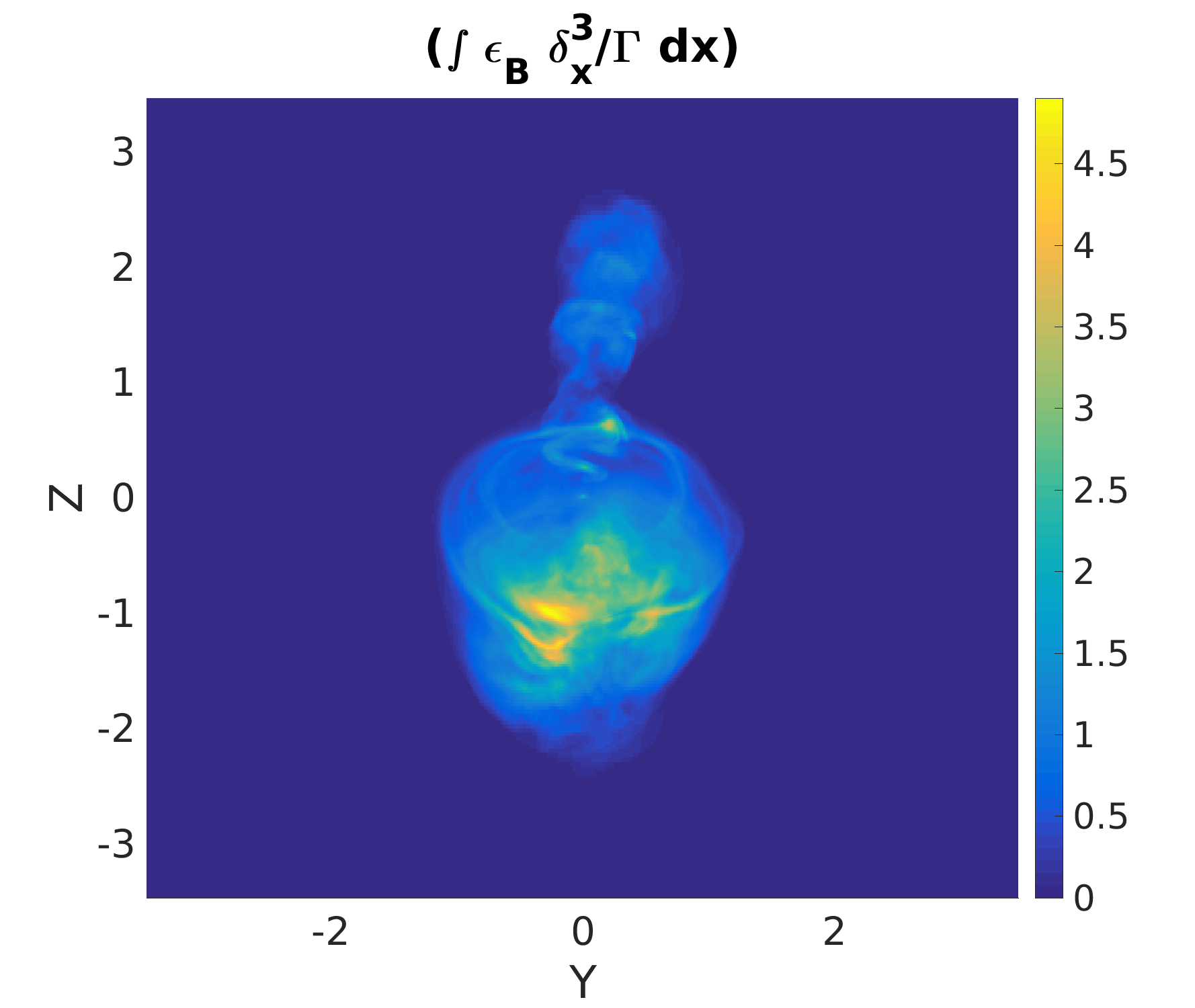}
\includegraphics[width=85mm,angle=-0]{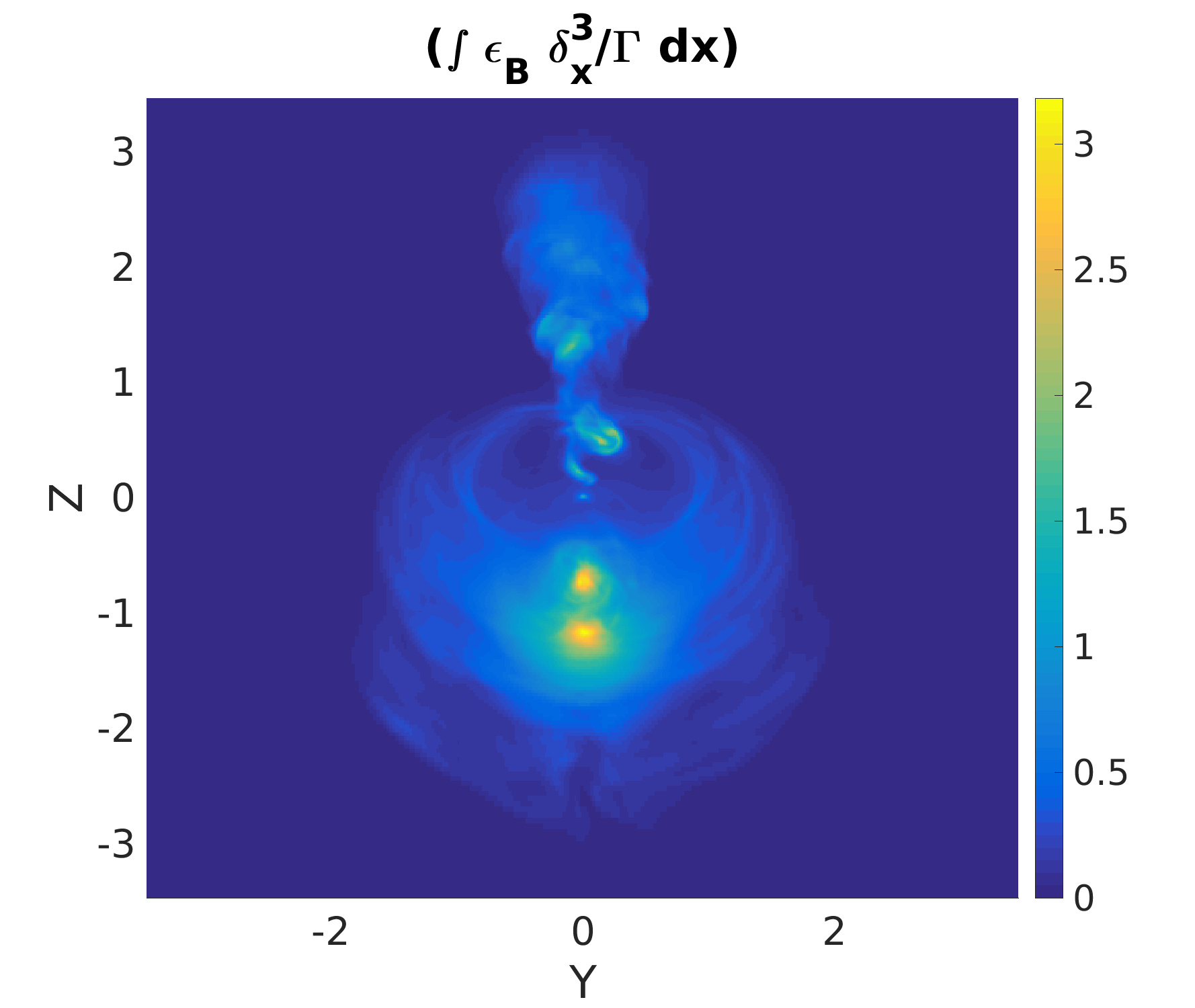}
\includegraphics[width=85mm,angle=-0]{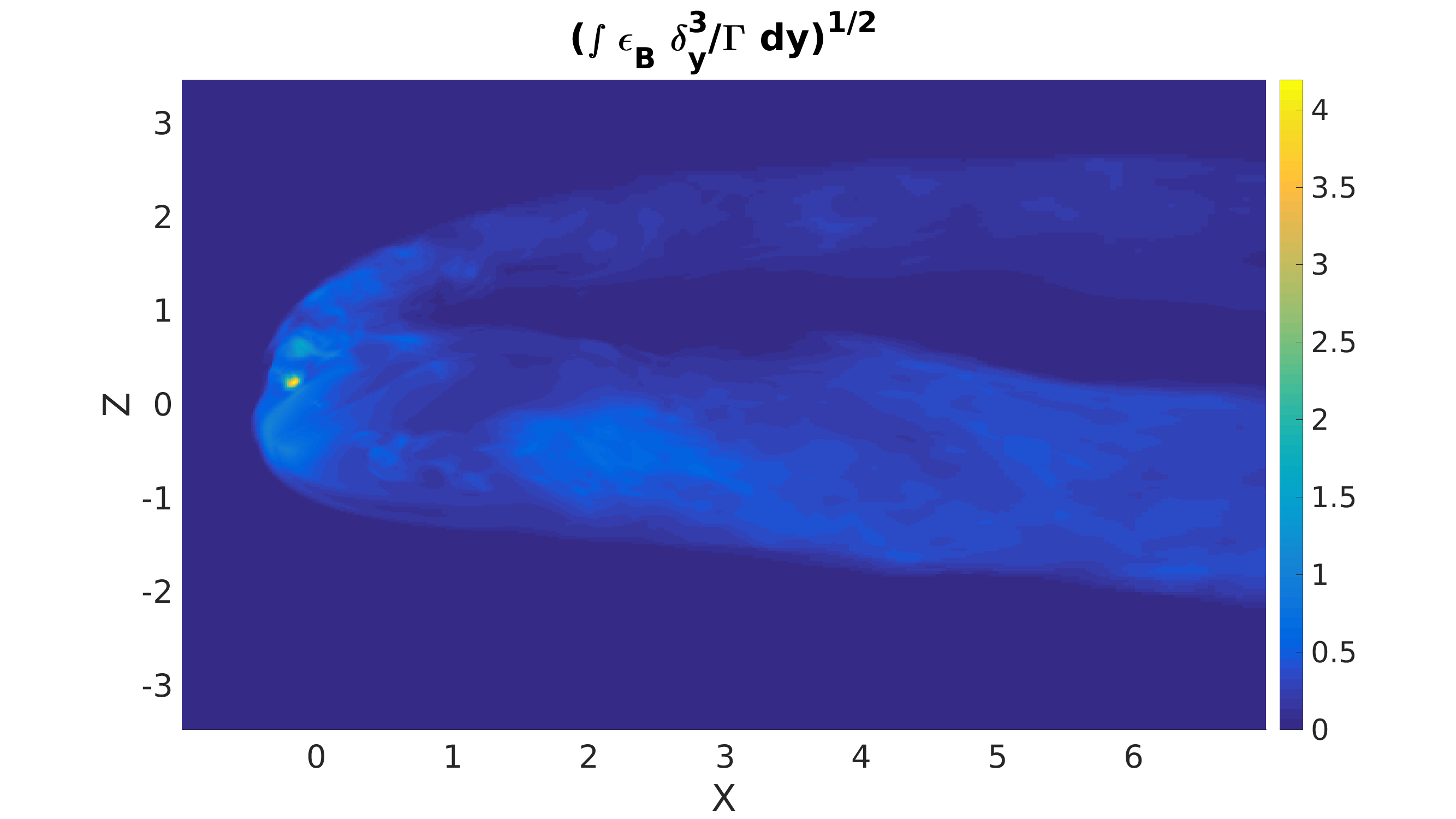}
\includegraphics[width=85mm,angle=-0]{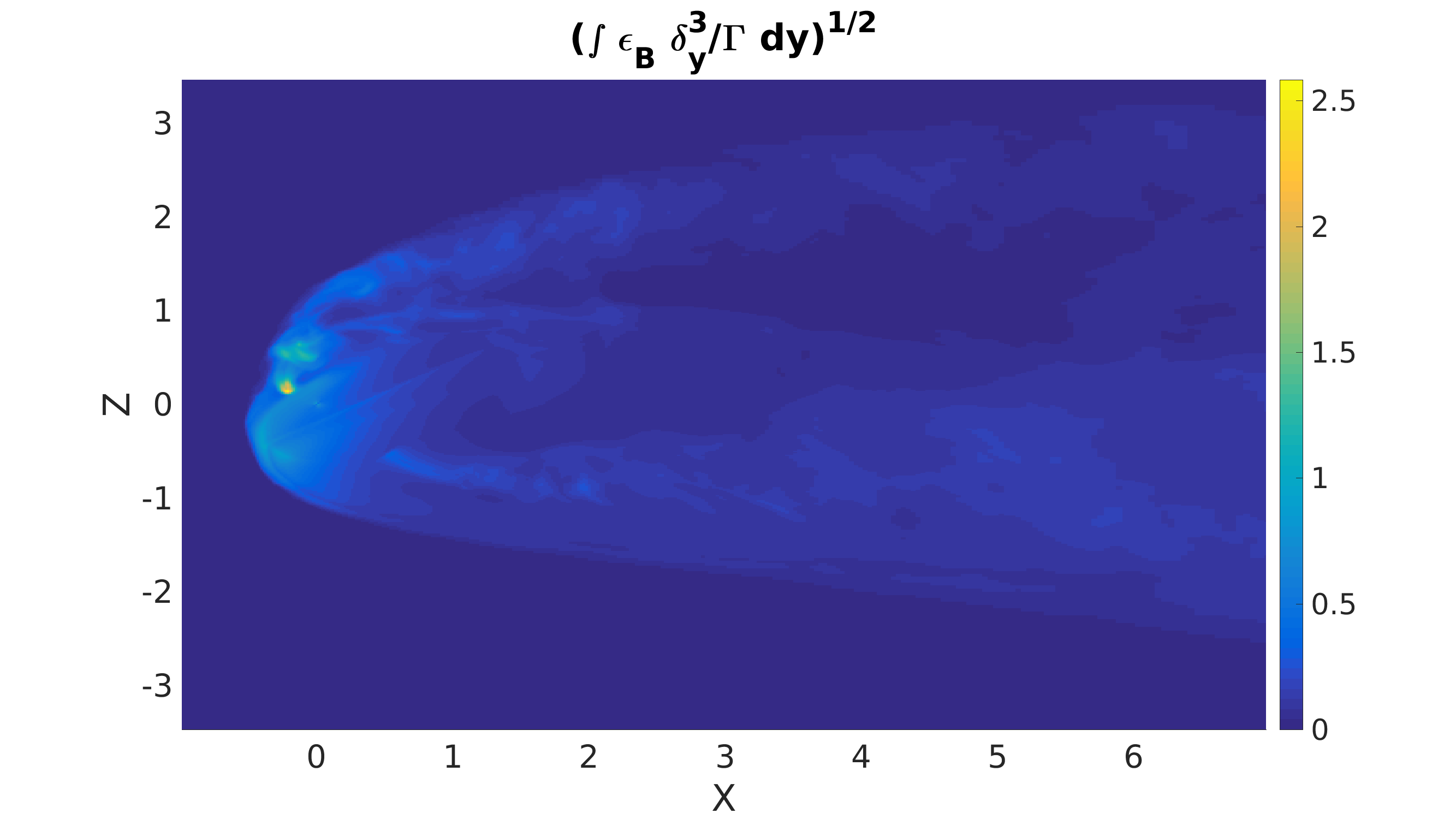}
\includegraphics[width=85mm,angle=-0]{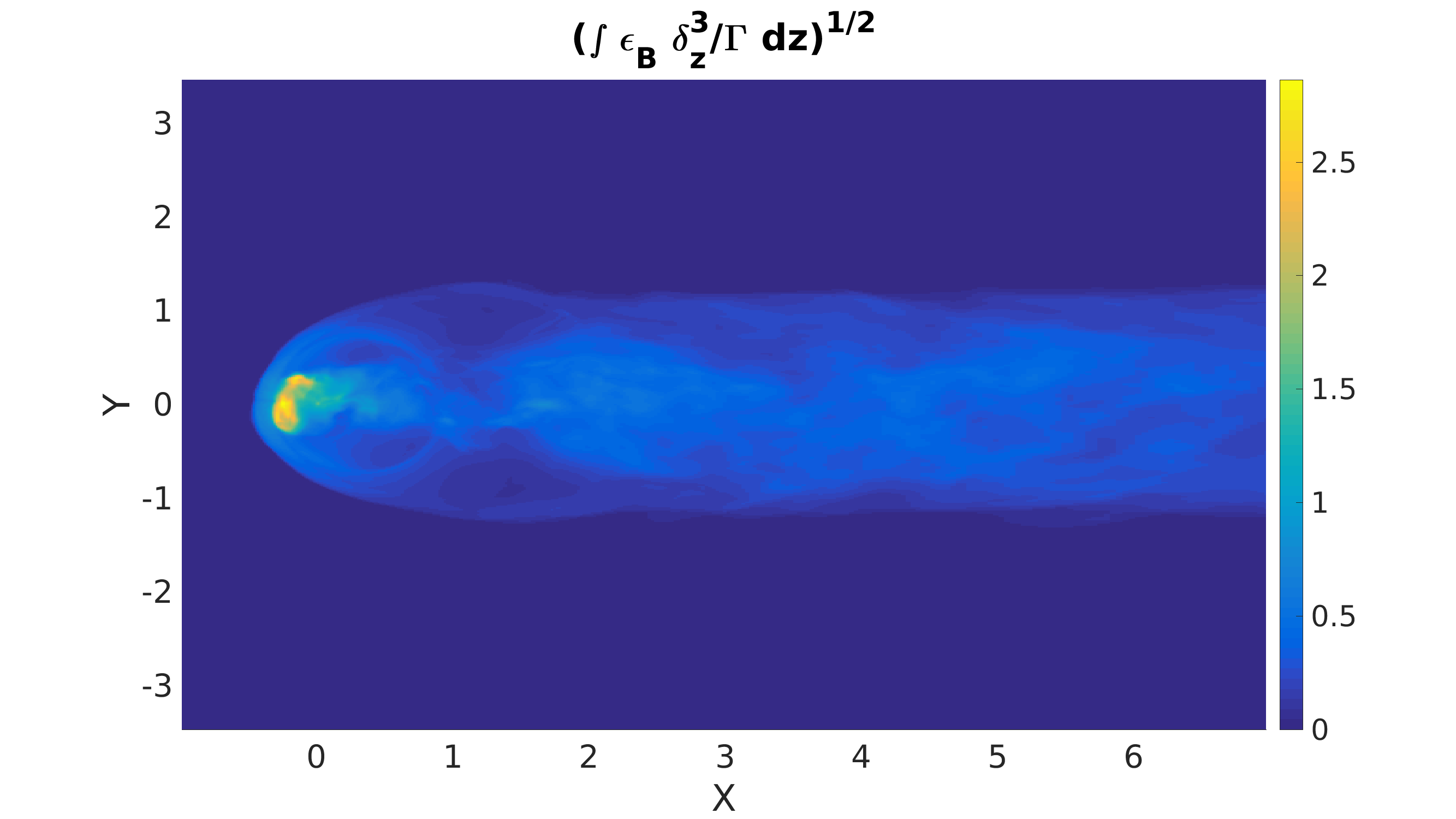}
\includegraphics[width=85mm,angle=-0]{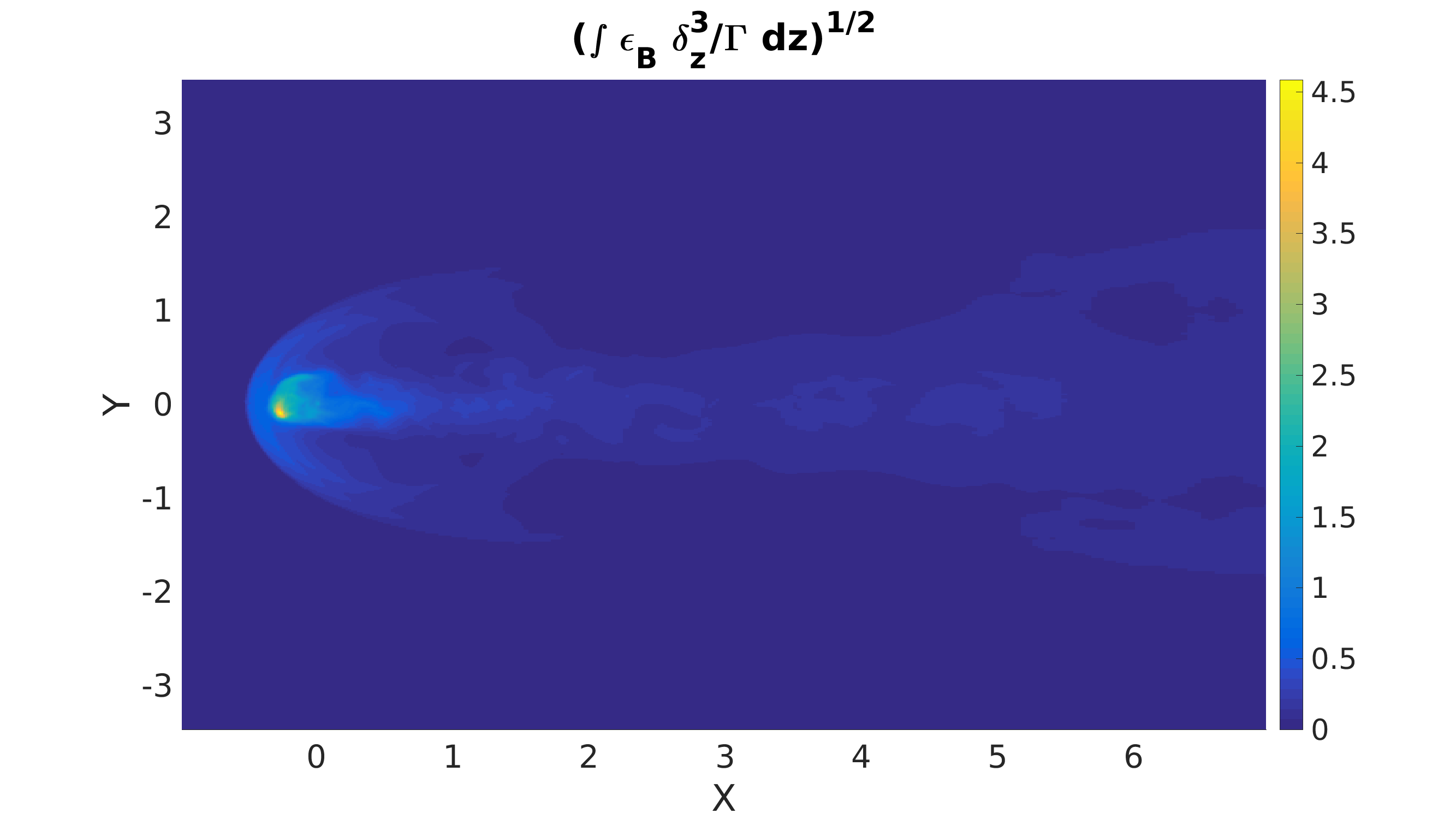}
\caption{Here we present the SYmapDopler emissivity map for model slow (left) and fast (right) projected along X axis (top), Y axis (middle), Z axis (bottom).}
\label{fig:Ebd}
\end{figure*}

We should note that the solution was calculated in the framework of ideal RMHD, which does not include reconnection processes. The reconnection presented in our simulations is the result of numerical viscosity and is sensitive to the spatial resolution of the computation grid. This will be examined further in a future work.

\subsection{Magnetic field geometries}
\label{s:geom}

One of the most surprising facts about kinetic jets is that they are highly asymmetric, extending far out on one side of the pulsar (see, e.g., Fig.~\ref{fig:lighthouse}). 
These asymmetries can be explained as the effects of asymmetric magnetic bottling at the incoming and outgoing parts of magnetic field lines, see Fig.~\ref{fig:sl0}. Qualitatively, the paths for escaping particles along incoming and outgoing parts of the \Bf\  lines are somewhat different due to intrinsic asymmetries of the PWN and orientation of the ISM field.

The PWN-ISM interaction depends both on the intrinsic properties of the PWN and the orientation of the ISM \Bf. To distinguish various intrinsic geometries we
introduce three distinct
orientations of pulsar spin axis relative to its direction motion in the ISM and the plane of the sky: (i) bullet geometry, where the pulsar rotation axis is parallel to the direction of motion and both are in the plane of the sky;
(ii) frisbee geometry, where the pulsar rotation axis is perpendicular to the direction of motion and lies in the plane of the sky;  (iii) cartwheel geometry, where the pulsar rotation axis is parallel to the direction of motion and is perpendicular to the plane of the sky \citep[for more details see ][]{BL17a}.  The frisbee and cartwheel configurations are physically the same but have different lines of sight.

If the ISM magnetic field lines are perpendicular to the direction of {the} pulsar motion and the pulsar has a frisbee or bullet orientation, then each of the ISM magnetic field lines has another magnetic field line in ISM which, after reconnection, will be equally likely to leak non-thermal particles from the pulsar wind zone. Such configurations will form symmetric kinetic jet-like structures. 
For asymmetric kinetic jets to form, the symmetry must be broken.  
This can happen by a mixed configuration between frisbee and bullet (i.e., if the pulsar spin axis is not perpendicular or parallel to the direction of motion), or if the magnetic field is not oriented perpendicular to the direction of motion.  Below we will model the former configuration.

\subsection{Initial setup}
\label{s:inc}

We start our simulation from a non-equilibrium configuration and evolve it up to a time at which 
quasi-stationary solutions will be settled. 
From the left edge $(X=-4)$ we inject the ISM with  Mach number $\mathcal{M}=85$ ($\mathcal{M}\equiv v_{\rm PSR}/c_{\rm ISM}$). 
We setup the ISM speed as 
$v_{\rm ISM}=0.0033 c$ (slow model) and $v_{\rm ISM}=0.1 c$ (fast model).
Unrealistically high ISM speed leads to drastic saving of computational resources, which is proportional to $\propto 1/v_{\rm ISM}$.
The density of the ISM was  adopted so that in the case of non-magnetized  spherical pulsar wind the bow shock will be formed at a position 
near $(-1,0,0)$, so our grid is scaled to stand-off distance $r_s$ (see Eq.~\ref{eq:rs}). The smaller value of the ISM speed value is more realistic and corresponds to {the} velocity of \psrA, the higher velocity corresponds to the ISM velocity of our another study  \citep[][]{BL17a}.

Here we use the pulsar wind setup described 
by \cite{2014MNRAS.438..278P,BL17a}.
A pulsar with radius 0.2 is placed at the point (0,0,0). The pulsar emits  unshocked magnetized pulsar
wind with a toroidal magnetic field which changes its polarity in the north and south hemispheres. 
The pulsar wind was injected with initial bulk Lorentz factor $\Gamma = 1.9$, magnetization $\sigma=1$ and Mach number 15.
We choose one orientation, frisbee-bullet, which is formed by a clockwise rotation of the frisbee configuration\footnote{In the frisbee orientation pulsar spin axis is parallel to Z-axis.} around 
Y axis by angle $\theta=\pi/4$ \citep[see for more details][]{BL17a} i.e. we set an angle of $\pi/4$ between spin axis and the pulsar velocity, see Fig.~\ref{fig:op}.

The ISM has a weak magnetization $\sigma_{ISM}=0.01$, formed by a uniform magnetic field, which is normal to the velocity vector (parallel to the ZY plane) and inclined by $\pi/4$ counterclockwise relative to the Z-axis.

\begin{figure*}
\includegraphics[width=0.53\linewidth]{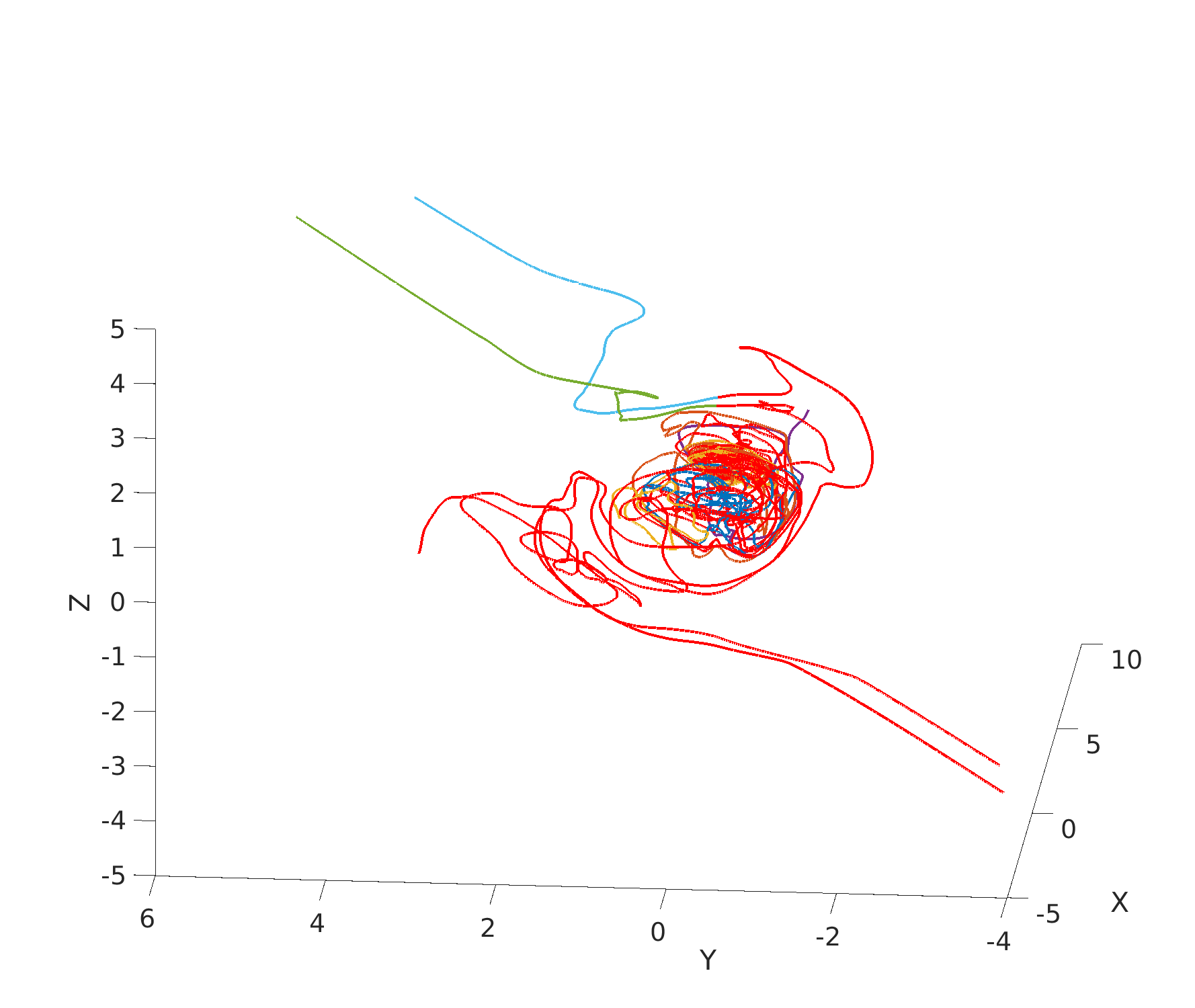}
\includegraphics[width=0.45\linewidth]{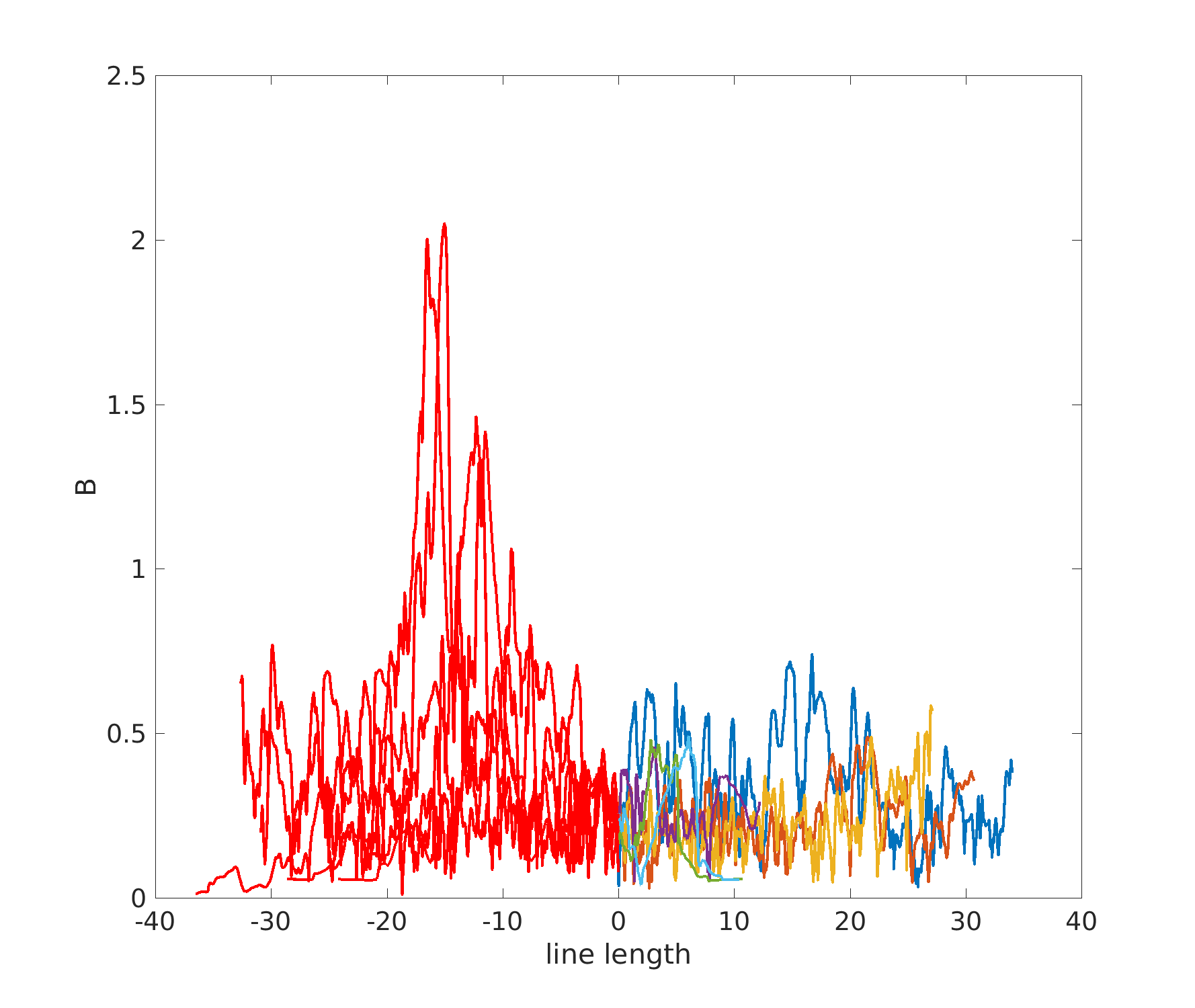}
\caption{Left Panel: Magnetic field streamlines for the slow model. The magnetic field lines enter the pulsar wind via reconnection at the boundary. We place magnetic field line starting points on the plane $Y=0$ and $X=[5,7]$ and $Z = [-1.2,0,1.2]$. 
Right Panel: value of \Bf\ along the \Bf\ lines. Note the left-right asymmetry (corresponding to the incoming-outgoing parts  of \Bf\ lines). In the particular example the left section (negative abscissa points) has much larger fluctuations of field strength.  This will lead to formation of magnetic bottles that would reflect a larger faction of PWN particles than the more smooth section on the opposite side (positive abscissa points), leading to asymmetric outflows.}
\label{fig:sl}
\end{figure*}

\section{Results}
\label{s:res}

The anisotropic ram pressure of the ISM forms a bow shock (see Fig.~\ref{fig:rho}). The pulsar wind is deflected to the direction opposite of the pulsar motion. 
Inside of the contact discontinuity the pulsar winds form two jet-like outflows.
The part of the front top outflow is deflected and turned down and back. This deflected flow has a magnetic field with an opposite twist direction compared to the magnetic field twist of the bottom back flow.
The pulsar wind forms an asymmetric structure which has a relatively simple jet-like structure on top with homogeneous twist; the bottom flow is stronger and has a more complicated magnetic field topology with the reverse magnetic field twist structure, see Fig.~(\ref{fig:sigma}). The magnetic field lines there are launched in both directions from starting points on a sphere of radius 1.5 placed at point (6,0,0).   

Comparing the slow and fast models we see generally similar solutions:  the formation two jet-like structures with simpler structures at the upper part and more sophisticated at lower parts.  
On the other hand, Fig.~(\ref{fig:rhoB}) shows significant differences in the number of magnetic field lines from the ISM which penetrate into the pulsar wind zone. This fact can be explained by differences in the ISM speed and 
instabilities, which trigger the reconnection and have more time to grow in the slow model. 

\subsection{Emissivity maps}

Here we create illustrative emissivity maps of two kinds, using the same approach as described in our previous work \citep{BL17a}.
In an ideal RMHD simulation, we have no direct information about non-thermal particle spectra and densities,
so we assume that in the shocks some fraction of thermal energy is transfered to NT particles. Also, here we  
neglect by-effects connected to NT particle spectra. 
The volume emissivity for Synchrotron radiation is proportional to the NT particle's density and inverse proportional cooling time, which  
is proportional to the co-moving magnetic field energy density (see Fig.~\ref{fig:Ebd}).

In both Figure \ref{fig:Ebd} we see the formation of asymmetric jet-like structures and fainter equatorial outflows. In 
the synchrotron maps, jets-like structures {appear} to be much more visible. If {the} pulsar moves towards us (upper panels), we see an asymmetric structure with a brighter bottom/back jet structure. If we view at the PWNe from {the} side (perpendicular to its direction of motion), we see the asymmetric jet-like structure. If we view the system from the ``top", jet-like structures merge and becomes bright head narrow tail structure.  In all cases, on the synchrotron maps, the equatorial flow is less visible. Morphologically, the fast and slow models are similar.

 As we can see above the both models have very similar morphologies and emissivity maps, so the results of \cite{BL17a} can be safely extrapolated to much slower pulsars.

\section{Dynamics and morphology of kinetic jets}
\label{s:applications}

\subsection{Magnetic field connection time}
The stand-off distance, Eq.~(\ref{eq:rs}),  is 
$r_s\approx 10^{16} \, n_{0}^{-1/2}\, \v_{p,8}^{-1} $~cm.
The travel time to the stand-off distance is given by
\be 
t_{s} = \frac{r_s}{v_p} = 4  n_{-1}^{-1/2}\, \v_{p,8}^{-2} \mbox{ yrs}
\label{ts}
\ee
We expect that $t_s$ is the typical time that any ISM field line remains connected to the pulsar wind.

For an ISM field of value $B$ and observed photon energy $\epsilon$, the required \Lf\ of the radiating particles is
\be 
\gamma_w \sim \sqrt{\frac{ m_e c \epsilon_{\gamma} }{e \hbar B}}
= 2 \times 10^7 \epsilon_{\gamma,\rm 1\, keV}^{1/2} B_{\rm 100 \mu G}^{-1/2}
\label{gammaw}
\ee
Such values can be expected in pulsar winds, \cf\ \cite{1984ApJ...283..694K}.
The corresponding synchrotron cooling time, 
\be 
\tau_c =\left(\frac{3}{2}\right)^{5/4}\left(\frac{ {\hbar } m_e^{5}c^{9}}{ e^{7} {\epsilon_\gamma }B^{3}}\right)^{1/2}=
60 \, \epsilon_{\gamma,\rm 1\, keV}^{-1/2} B_{\rm 100 \mu G}^{-3/2}\; {\rm yrs},
\ee
is comparable, but somewhat larger than the travel time (\ref{ts}). Thus, we do not expect any spectral evolution along the kinetic jets.

\subsection{Jet anisotropy}

One of the most surprising fact about kinetic jets is that often  they are highly asymmetric, extending far out on the one side of the pulsar, \eg, Fig. \ref{fig:lighthouse}.
 Below we discuss two possibilities for this asymmetry - Doppler boosting and asymmetric magnetic bottles.
After rejecting the Doppler boosting possibility, \S \ref{boost}, 
we explain these asymmetries due to effects of asymmetric  magnetic bottling, see \S \ref{bottle}.

\subsubsection{Doppler boosting?}
\label{boost}

Can  the brightness difference between the X-ray structures propagating out of the Lighthouse Nebulae be due to Doppler boosting/de-boosting?

The luminosity of the continuous jet with Dopler boosting can be written as \citep{1997ApJ...484..108S}
\begin{equation}
L_{\rm obs}= \frac{\delta^3}{\Gamma_{\rm jet}}L_{\rm emitted},
\label{eq:lboost}
\end{equation}
where $\delta = 1/[\Gamma_{\rm jet}(1-\beta\cos\theta)]$ is the Dolper factor, $\Gamma_{\rm jet}=1/\sqrt[]{1-\beta^2}$ is the Lorentz factor of the jet particles and $\beta=v_{\rm jet}/c$ is bulk flow velocity of the jet. If we have two symmetrical, jets the relative brightness $A\equiv (L_{\rm obs}'/L_{\rm obs}'')^{1/3}\ge 1$, where $''$ and $'$ denote the "approaching" and "receding" outflows, then from Eq.~\ref{eq:lboost} we obtain the following expression
\begin{equation}
\beta\cos\theta = \frac{A-1}{A+1}\;.
\label{eq:bth}
\end{equation}
Thus, we can place a very modest restriction on the bulk Lorentz factor and mildly-relativistic outflow with $\beta>0.4$ or $\Gamma>1.1$ to explain the brightness difference $A\sim 2$ or $L_{\rm obs}'/L_{\rm obs}''\sim 10$. 

On other hand, large Lorentz factors will leads us to $\beta\approx 1$ and a viewing angle cosine of $\cos\theta \approx (A-1)/(A+1)$. The Doppler factor will take the following form  $\delta \approx (A+1)/2\Gamma_{\rm jet}$. 
The Doppler boosting will require a jet luminosity on the level 
\begin{equation}
L_{\rm emitted} \approx \frac{8\Gamma_{\rm jet}^4}{(A+1)^3}L_{\rm obs}.
\label{eq:limin}
\end{equation}
Taking into account the total energy budget of the PWNe ($L_{\rm emitted}\ll\dot{E}$), we can find an upper limit on the bulk jet's Lorentz factor as
\begin{equation}
\Gamma_{\rm jet}\ll\left(\frac{(A+1)^3}{8}\frac{\dot{E}}{L_{\rm obs}}\right)^{1/4}\sim 5 \dot{E}_{36}^{1/4}L_{\rm obs, 33.5}^{-1/4}.
\label{eq:gmax}
\end{equation}


Hence, we conclude that the brightness difference between the oppositely propagating highly relativistic jets can be not only due to Doppler boosting/de-boosting.

\subsubsection{Jet anisotropy - non-symmetric magnetic bottles}
\label{bottle}

As the external \Bf\ connects to the pulsar wind, we expect that 
the incoming and outgoing parts of magnetic field lines will have different structure, see Figs. \ref{fig:sl0} - \ref{fig:sl} for qualitative description.  The paths for escaping particles along incoming and outgoing parts of the \Bf\  lines are somewhat different due to intrinsic asymmetries of the PWN and orientation of the ISM field. Fluctuations of the \Bf\ will generate in the collisionless plasma magnetic bottles that will partially prevent the particles from escaping.

Let us consider isotropic injection of relativistic particles onto an open field line (e.g., near the point $x=6$ in Fig. \ref{fig:sl}). As particles propagate along the field they experience partial reflection at magnetic bottles due to the requirement of conserved first adiabatic invariant $\propto \sin^2 \psi/R$, where $\psi$ is the particle pitch angle and  $R\equiv {B}/{B_0} \geq 1$ is the  mirror ratio. In the case of no scattering, the flux of particles that pass through depends on the highest value of the  mirror ratio along a given field line.

The conservation of the first adiabatic invariant in a collisionless plasma leads to the following gyration-averaged equation for the distribution function
\citep{1969lhea.conf..111R,Kulsrud}
\be 
\mu \partial_z f - \frac{1}{2} \partial_z \ln B \, (1-\mu^2)\partial_\mu f=0
\label{drift}
\ee
where $\mu =\cos \psi$, $\psi$ is the pitch angle. Equation
(\ref{drift}) can be integrated along particles' trajectories $(1-\mu^2)/R=$constant to give
\be 
f= f\left( (1-\mu^2)/R \right)=f_0(\mu_0),
\label{f}
\ee
where $f_0(\mu_0)$ is a given pitch angle distribution at $R=1$.

Suppose a magnetic mirror forms between the internal PWN plasma with isotropic distribution and the ISM. The distribution of the escaping particles in the ISM then remains isotropic, while 
 the fraction of particles that pass through the mirror  is 
\be
p_{esc}= 1-\sqrt{1-\frac{1}{R}} \approx \frac{1}{2R}
\ee
where the last relation assumes $R\gg 1$ (at the same time the escaping flux is $1/R$). For multiple mirrors and without any scattering, it is the maximal $R$ that determines the escaping fraction.5

Thus, we expect that if two escaping trajectories (at the entry/exit points of the \Bf\ lines) encounter different magnetic bottles, the ratio of the escaping fraction  will be of the order of the ratio of the maximal mirror ratios $R$. 

For example, according to Fig. \ref{fig:sl}, the average field near the origin is $B\sim 0.2$ (in arbitrary units), while the maximal field at negative line coordinates is $B\sim 2$. Thus, $R_{\rm PWN} \sim 10$ and $p_{esc} \approx 0.05$ -- only $5\%$ of injected particles will escape. On the other hand, at  positive line coordinates, the maximal field is $B\sim 0.5$, which gives $R \sim 2.5$ and $p_{esc} \approx 0.22$ - approximately four times larger.   

In the case of the Lighthouse PWN, \citet{ppba16} find that the ratio of jet-counter jet surface brightness is much higher, $\sim 40$. This high ratio can be due, \eg, to special configurations of the fields not captured in our simulation (due to high computational costs we were not able to explore in detail such specific configurations). Alternatively, pitch angle scattering between the multiple magnetic bottles may lead to a higher jet-counter jet  brightness ratio. (For example, if particles are isotropized between each bottle, the total escaping fraction will be the product of the escaping fractions from each bottle.)

\subsection{Structure of kinetic jets away from the PWN: magnetic bottles--anti-bottles?}
Next, let us discuss  what determines the morphology of the kinetic jets away from the PWN. For example, in the case of the Lighthouse nebula, Fig.\ref{fig:lighthouse}, the kinetic jets consist of several elongated features - what processes give rise to these features? Importantly, the elongated features show nearly constant emissivity along their length. This places interesting constraints on their nature as we discuss next. 

\begin{figure*}
\includegraphics[width=0.45\linewidth]{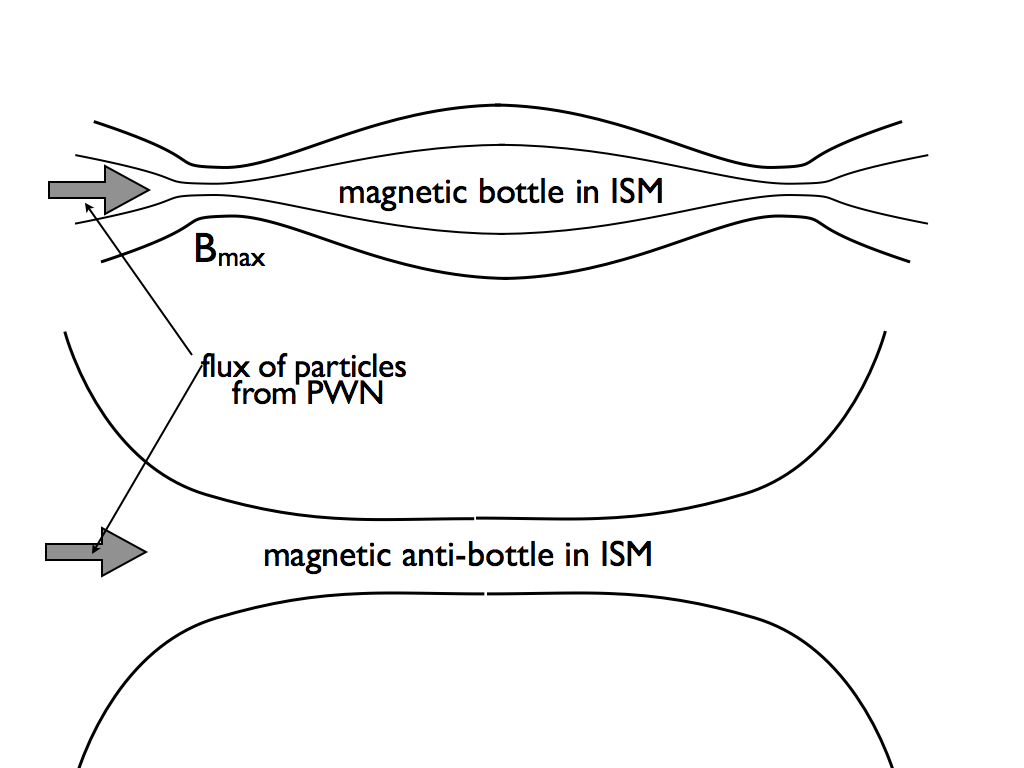}
\includegraphics[width=0.45\linewidth]{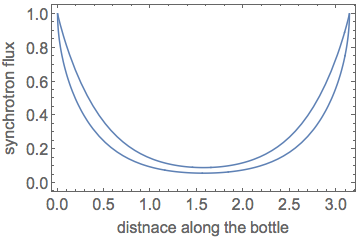}
\caption{Synchrotron emission from magnetic bottle and anti-bottle in the ISM. A kinetic flux of PWN leptons propagates along \Bf\ lines and encounters a magnetic bottle/anti-bottle (left panel). The corresponding synchrotron signal along the bottle integrated over the cross-section (right panel). Two curves correspond to isotropic injection (lower curve) and parallel beam with $f\propto \mu_0^2$ (upper curve). The synchrotron profile of a magnetic bottle  should have peaks at the location of the bottle necks, contrary to observations. In the case of anti-bottle the flux should peak in the middle.}
\label{fig:bottle-ISM}
\end{figure*}

One possibility to produce disconnected elongated features is with magnetic bottles in the ISM, see Fig. \ref{fig:bottle-ISM}. Consider a beam of particles with angular distribution $f \propto \mu_0^{\alpha}$ at bottle neck (where magnetic field is maximal, $B_{max}$). As the beam propagates into the regions of smaller \Bf\ with $R=B/B_{max}< 1$ the pitch angle of the particles decreases; at any given $R$ the maximal pitch angle correspond to  $\mu _{min} = \sqrt{1-R}$. The synchrotron emission integrated over the bottle cross-section is then
\be 
\propto R^2 \frac{1}{R} \int_{\mu _{min}}^{1} f_0(\mu_0(\mu, R)) (1-\mu^2)
\ee
where the terms $R^2$ and  $(1-\mu^2)$ account for synchrotron luminosity $\propto B^2 \sin^2 \psi$, 
the term $1/R$ accounts for changing cross-section of the bottle  $S$ (magnetic flux is conserved, $S B=$constant), in the distribution function the initial pitch angle $\mu_0$ should be expressed in terms of current $\mu$ and mirror ration $R$.

As an example, let us consider two cases of isotropic distribution $\alpha=0$ and parallel beam $\alpha=2$. Let the magnetic bottle have the minimal \Bf\ two times smaller than in the bottle neck, \eg, 
$R=1- (1/2) \sin z$, here z is coordinate along magnetic field line. The resulting synchrotron profile has peaks near the bottles' necks, contrary to observations. We conclude that the morphology of the kinetic jets is not due to magnetic bottles in the ISM.

Analogously, in the case of anti-bottle, the synchrotron flux  peaks in the region of highest \Bf\ (approximately by the mirror ratio  $R>1$, defined with respect to some fiducial value. For long bottles with nearly constant cross-section this will produce nearly constant synchrotron luminosity.   


In conclusion, both ballistically moving kinetic jets as well as magnetic  anti-bottles in the ISM may reproduce the morphological feature. The key distinction is the temporal changes: in the case of  ballistically moving jets we expect to see very quick morphological changes. In another case, we predict changes in the features on the time scales of a $\sim$ decade, see Eq. (\ref{ts}).

\subsection{Jet-induced kink instability}
Finally, let us comment on a suggestion by 
\cite{2014A&A...562A.122P} that the features of the kinetic jet arise due to kink instabilities produced by the streaming of escaping particles. Pulsars produced charge separated flows  with typical currents of the order of
\be
I \sim \sqrt{ c \dot{E}}.
\ee
If a fraction $\eta_{esc}\ll 1 $  of  this current  escapes  through a hole  of  the  typical  size
 $r_{esc}\sim\sqrt{\eta_{esc}}\, r_s$,  this will produce an induced (toroidal)  magnetic field in the ISM  order of
$B_{ind}\sim 2I_{esc}/c r_{esc} \sim v_{p} \sqrt{4\pi\eta_{esc}  \rho_{\rm ISM}}$. 
 The ratio of the induced magnetic field  to the ISM field  is then
 \be
 \frac{B_{ind} }{B_{ISM}} \sim \sqrt{\frac{\eta_{esc}}{\beta_{\rm ISM}}} M_{ISM} 
 \ee
 where $M_{ISM} \gg 1$  is  the  Mach number  of the pulsar  with respect to the ISM  sound speed,  and $\beta_{ISM}=p_m/p_g \leq 1$  is  the ISM  beta parameter.  This ratio can be,  under certain circumstances,  larger than  unity,   so that the magnetic field  induced by the escaping charge-separated flows  can indeed  distort the ISM  fields considerably.

\section{Application}

\subsection{The Lighthouse PWN (\psrA{})}

In this system a fairly energetic pulsar, PSR J1101--6101 (spin-down power $\dot{E} = 1.36 \times 10^{36}$ erg s$^{-1}$, distance $\sim 7$ kpc), 
produces an emission feature extending $11$ parsecs with a total $X$-ray luminosity $L \sim 3 \times 10^{33}$ erg s$^{-1}$  \citep{ppba16}.
The kinetic-jet light crossing time $t_{\rm lc}= D_L/c\approx 40 \;\rm yr$ can be close to the NT particle propagation time.
Thus, we do not expect spectral evolution along the kinetic jet, which is consistent with the roughly constant spectrum $\Gamma\approx1.6$ reported by \citep{ppba16}.

In the case of the Lighthouse PWN, the X-ray luminosity of the kinetic jet is 
$L \sim 3 \times 10^{33}$ erg s$^{-1}$ (for a distance of 7 kpc). This then requires 
\be 
N_{em} \approx \frac{ \tau_c  L}{\gamma_w m_e c^2} \approx 2\times10^{41}
\ee
emitting particles. 

Let us assume that after the termination shock the pulsar wind has magnetization $\sigma_w \sim 1$  
and bulk \Lf\ $\gamma_w$ given by Eq. (\ref{gammaw}),
so that spin-down power can be estimated as $\dot{E}= \gamma_w m_e c^2 \dot{N}$,  where $\dot{N}$ is number of pairs produced by the pulsar per second. 
The number of particles that a pulsar produces during time $t_s$ is then $N_p = \dot{N} t_{s}$.
Thus, the fraction 
\be 
\frac{N_{em}}{N_p} = \frac{\tau_c  L}{\dot{E} t_{s}}\approx \frac{L n_{-1}^{1/2}\, \v_{p,8}^{2}}{\dot{E}\epsilon_{\gamma,\rm 1\, keV}^{1/2} B_{\rm 100 \mu G}^{3/2}}\ll 1
\label{eq:NemNp}
\ee 
is required to preserve number of NT particles. Now we can derive a lower limit for the ISM magnetic field strength as 
\begin{equation}
B_{\rm 100 \mu G}\gg 0.02 \left(\frac{L_{33.5}^2n_{-1}\v_{p,8}^{4}}{\dot{E}_{36}^2 \epsilon_{\gamma,\rm 1 keV}}\right)^{1/3}
\label{eq:Bmin}
\end{equation}
Given the uncertainties in the parameters, $B_{\rm 100 \mu G}> 0.1$ is a reasonable estimation. 
In other words, the formation of kinetic jets requires not only the reconnection of the pulsar wind magnetic field with the ISM field, but also a high ISM magnetic field strength.  

The estimate of the connection  time (\ref{ts}) fits well with the linear size of the emission feature. {Although} the total size is $11$ pc, the morphology of the feature (see Fig. \ref{fig:lighthouse}) shows two separate seemingly-disjointed components, each roughly half the size of the jet, $\sim 5$ pc. This gives an excellent correspondence to the the connection  time (\ref{ts}) of $\sim 15$ yrs.

\subsection{G327.1--1.1 and MSH 11--62: ``Snail eyes'' Morphology}

\begin{figure*}
\includegraphics[width=0.45\linewidth]{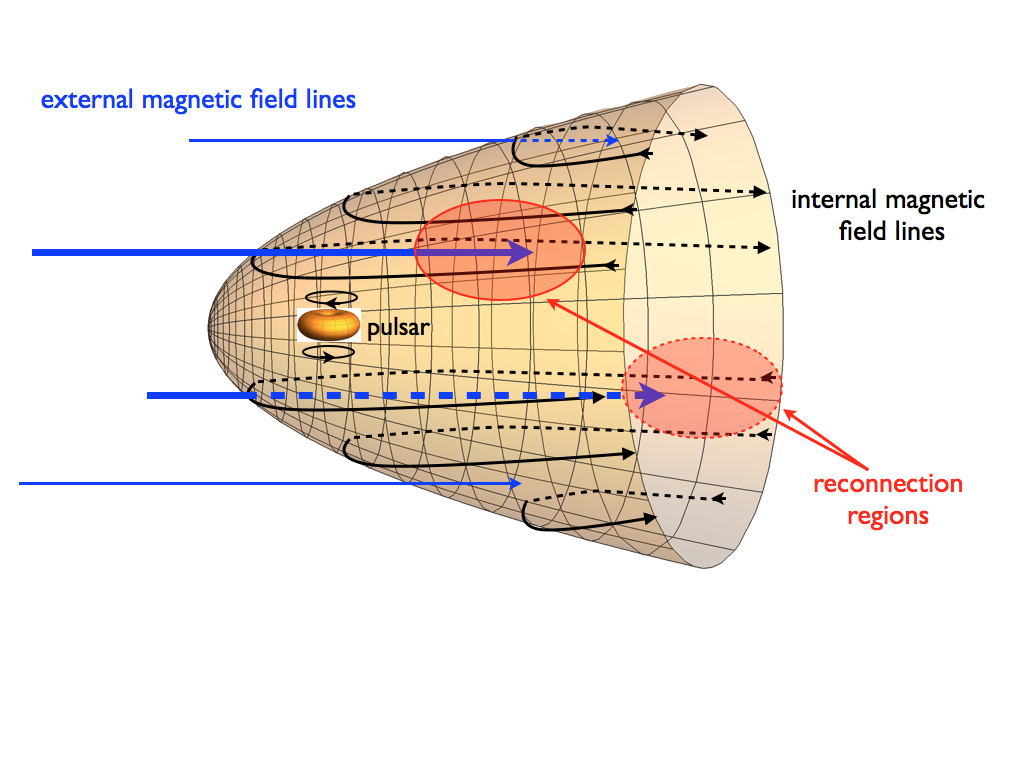}
\includegraphics[width=0.45\linewidth]{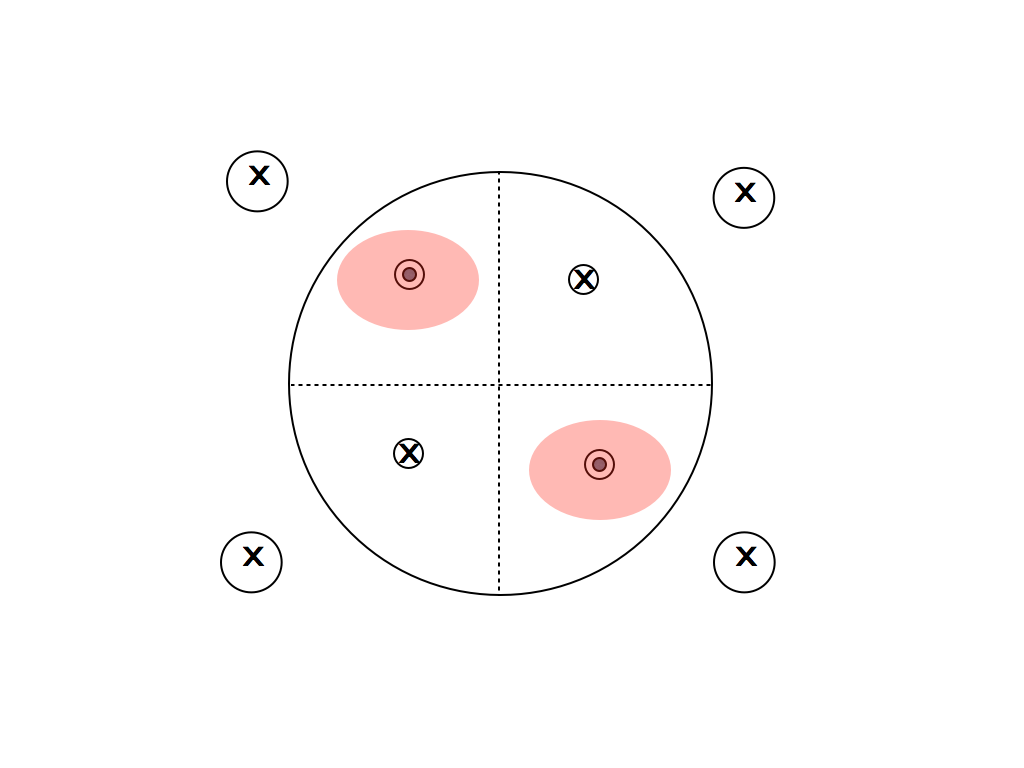}
\caption{Qualitative explanation of ``snail eyes'' morphology. In the frisebe/cartwheel geometry the internal \Bf\  changes direction in the four quadrants. (The internal magnetic field is depicted with black lines  - field lines in the forefront are solid, field lines in the rear are dashed.)
External \Bfs\ lines are parallel to the velocity vector. In the two opposite quadrants the external fields are counteraligned with the internal fields (thick blue lines), resulting in   reconnection of external and internal fields (red regions). Left panel: side view, right panel: head-on view.}
\label{fig:snail_sk}
\end{figure*}

In the case of a magnetic field aligned with the direction of {the} pulsar motion, the most probable reconnection {sites} will appear in diagonally placed quadrants relative to pulsar equatorial plane.  In the Fig.~\ref{fig:snail_sk} we {provide} a schematic sketch of the magnetic field topology formed in the frisbee  configuration (see Fig.~\ref{fig:rhoB} and the magnetic field topology in the work of \citealt{BL17a}). The red ellipses indicate the most probable reconnection zones. Such reconnection zones can form foot-points for the formation of double kinetic jets as seen in the cases of G327.1--1.1 and MSH 11--62.

The symmetry of the jets (both structure and brightness) of G327.1--1.1 (the Snail PWN) and MSH 11--62 indicate that the pulsars' motions are symmetric with respect to the magnetic fields ahead of the bow shock (i.e., they are aligned with the directions of pulsar velocity), and that reconnection is occurring in opposite sides of the bow shocks, in the manner illustrated in Figure \ref{fig:snail_sk}.  Indeed, radio polarimetry measurements have revealed that the magnetic field in the Snail PWN bow shock is aligned with the pulsar's direction of motion and the jets (see Figure 7 of \citealt{ma16}), and that the magnetic field both in and ahead of the bow shock are aligned with the jets and {the} inferred direction of motion of the pulsar in MSH 11--62 (see Figure 3 of \citealt{roger1986}), thus providing evidence of the parallel reconnection scenario.

\section{Conclusions}
\label{s:conc}

In this paper we advance a model of extended emission features observed in some bow shock PWNe as  kinetic flows of high energy particles that escaped pulsar winds due to reconnection between the wind and the ISM \Bfs, confirming the model of  \cite{ban08}. As the particles escape the PWN, they propagate along ISM \Bfs\ producing synchrotron emission that highlights the structure of the ISM \Bf.

We expect that the geometrical properties in each particular case -- inclination the pulsar magnetic moment to the  spin axis, the direction of pulsar motion and the orientation of the external \Bf, as well as magnetic structures in the ISM --  can lead to a  variety of shapes of the kinetic jets. For example, 
the strong asymmetry of the kinetic jets in case of the Lighthouse and the Guitar nebulae can arise from asymmetric  structures of  the reconnecting  magnetic filed lines, while the double kinetic jets (``snail eyes" morphologies) can result in cases when a frisbee/cartwheel-type PWN propagates along the external \Bf. 

We argue that the appearance of the kinetic jets strongly depends on the structure of the \Bf\ in the ISM, and, thus, can be used to probe  \Bfs\ in the ISM: kinetic jets highlight the high magnetic regions (anti-bottles) in the ISM 



{Finally, an} important prediction of the model is that the morphology of the extended features should change on time scale (\ref{ts}), about a decade.

\section*{Acknowledgments}
We appreciate to Oleg  Kargaltsev and Rino Bandiera for useful discussions.
The calculations were carried out in the CFCA cluster of National Astronomical Observatory of Japan.
We thank the {\it PLUTO} team for the possibility to use the {\it PLUTO} code and for technical support. 
The visualization of the results performed in the VisIt package \citep{HPV:VisIt}. 
This work had been supported by   NSF  grant AST-1306672, DoE grant DE-SC0016369 and NASA grant 80NSSC17K0757.


\end{document}